# Modeling Oligarchs' Campaign Donations and Ideological Preferences with Simulated Agent-Based Spatial Elections

Mason Wright [1]& Pratim Sengupta[2]
Mind, Matter & Media Lab
Vanderbilt University – Peabody College

## Abstract

In this paper, we investigate the interactions among oligarchs, political parties, and voters using an agent-based modeling approach. We introduce the OLIGO model, which is based on the spatial model of democracy, where voters have positions in a policy space and vote for the party that appears closest to them, and parties move in policy space to seek more votes. We extend the existing literature on agent-based models of political economy in the following manner: (1) by introducing a new class of agents – oligarchs – that represent leaders of firms in a common industry who lobby for beneficial subsidies through campaign donations; and (2) by investigating the effects of ideological preferences of the oligarchs on legislative action. We test hypotheses from the literature in political economics on the behavior of oligarchs and political parties as they interact, under conditions of imperfect information and bounded rationality. Our key results indicate that (1) oligarchs tend to donate less to political campaigns when the parties are more resistant to changing their policies, or when voters are more informed; and (2) if Oligarchs donate to parties based on a combination of ideological and profit motivations, Oligarchs will tend to donate at a lower equilibrium level, due to the influence of lost profits. We validate these outcomes via comparisons to real world polling data on changes in party support over time.

## Introduction

[1] The author is currently a PhD student in Computer Science at University of Michigan. He can be reached at Masondw@umich.edu.

[2] The author is an assistant professor at Vanderbilt University, and director of the Mind, Matter & Media Lab. He can be reached at Pratim.sengupta@vanderbilt.edu

Oligarchs are defined in the political economics literature as leaders of firms in a common industry that has barriers to entry (Stigler 1964; Acemoglu 2008). The term *oligarch* refers to a broad class of individuals – broadly speaking, members of "interest groups" (Downs 1957) or "pressure groups" (Rubin 1975) for leaders of firms in an industry with barriers to entry, which lobby the government for legislative action to favor their common interest. Examples of oligarchs include "economic groups," such as industries and corporations (Stigler 1971), and "privileged groups" of individuals with a common interest, such as trade associations (Olson 1965). Oligarchs in an industry may tacitly collude, without directly communicating with each other, seeking to “capture” legislation that benefits them (Stigler 1964).

Pressure groups may be able to capture legislation through their influence, even in cases where the legislation benefits the oligarchs but has a cost to the rest of society (Stigler 1971). Special interest groups can obtain benefits, at a cost to the voters, because: (1) voters have imperfect information about politics and may not know the special interest benefits exist; and (2) political parties are willing to risk alienating voters by favoring special interests, if the parties are rewarded with campaign contributions that make up for the votes lost (Rubin 1975; Stigler 1971).

Empirical research shows that oligarch spending influences political outcomes in the US. Gilens and Page (2014) found that economic elites and business-oriented interest groups both had independent influences on policy outcomes, but the median voter did not. In a related study, Baumgartner, Berry, Hojnacki, Leech, and Kimball (2009) concluded that trade associations (e.g., the American Petroleum Institute) and corporations are more likely to mobilize effectively on an issue of particular interest to them than mass-based interest groups (e.g., labor unions, American Association of Retired Persons). Based on these studies, Gilens and Page (2014) have fittingly claimed that American democracy is subject to "economic elite domination" and "biased pluralism," where biased pluralism means that business-oriented interest groups have a disproportionately large effect on policy outcomes.

In this paper, we investigate the interactions among oligarchs, political parties, and voters using an agent-based modeling approach. We introduce the OLIGO model, which is based on the spatial model of democracy (Kollman, Miller, & Page 1992; Laver 2005), where voters have positions in a policy space and vote for the party that appears closest to them, and parties move in policy space to seek more votes. We extend previous work by Laver in the following manner: (1) by introducing a new class of agents—oligarchs—that represent leaders of firms in a common industry who lobby for beneficial subsidies through campaign donations; and (2) by investigating the effects of ideological preferences of the oligarchs on legislative action.

In doing so, we extend the literature by contributing a new model of special interest groups’ campaign donations that combines elements from agent-based models of voter-party interaction (Kollman, Miller, & Page 1992; Laver 2005), with elements from economic models of special interest group behavior (Rubin 1975; Bardhan & Mookherjee 2000). We verify the model's correctness with a series of basic hypothesis tests, such as to show that Parties in our model adopt a mean ideological position near the center of policy space. Furthermore, we validate the model's fit to real world data by comparing changes over time in Party support in our model with 10 years of nationwide polling data on party support in the US. We also provide results from novel simulation experiments where oligarchs have ideological preferences, and show that oligarchs have a lower equilibrium level of political action and influence, when they allow ideological rather than profit motivations to guide their campaign donations. That is, if oligarchs choose which party to support based on ideological goals instead of expected profits, and adjust their donation levels based strictly on subsequent profits, then equilibrium donation levels will be lower.

Our work accomplishes the following goals: (1) We show that the simple rules agents follow in our model are sufficient to capture much of the complex dynamics of this politico-economic system; (2) we verify and validate the results from prior studies that used analytic methods, using an alternative, agent-based modeling method; (3) we propose a novel finding regarding the nature of the equilibrium state of campaign donations of oligarchs with ideological preferences; and (4) we derive support for the claim that the OLIGO model is a useful test environment for novel hypotheses about oligarchs' campaign donation behavior.

# Politico-Economic Models and Agent-Based Models (ABMs)

## Prior work on Politico-Economic Models using ABM

Political scientists have studied oligarch behavior, or special interest group behavior in general, through building "politico-economic models" (Frey & Schneider 1975). The economic theory of politics provides a theoretical foundation for these models, in the idea that politics can be understood as a competition for influence among agents who seek to increase their own individual utilities. In a politico-economic model, parties compete for votes, and special interest groups seek to increase profits by competing for political influence and beneficial legislation. During a run cycle of a typical politico-economic model, a "popularity function" executes, where voters vote for the party with the most favorable policy position, and a "reaction function" executes, where parties change their policy positions to seek more votes in the next cycle (Frey & Schneider 1975). Examples of such models include Rubin (1975), Aumann and Kurz (1977), and Brock and Magee (1978).

More recently, Grossman and Helpman (1996) presented a model where two political parties compete for votes in a policy space with two issue dimensions. In this model, some voters are “informed” and others are “uninformed,” where informed voters have more accurate knowledge about parties' positions in policy space and are less susceptible to campaign advertising. Bardhan and Mookherjee (2000) developed a model that is especially similar to the model we will present here, although Bardhan and Mookherjee's model is analytic rather than agent-based. Bardhan and Mookherjee's model has two political parties competing for votes in an environment with one lobby group and voters who are either informed or uninformed. Informed voters vote for the party that maximizes their utility, while uninformed voters vote for whichever party spends more on campaigning.

The dynamics among oligarchs, voters, and political parties make up a complex, emergent system, where aggregate outcomes, such as the mean tax rate over time, emerge from interactions between individual agents. For example, parties do not follow any explicit rule that tells them to remain near the center of policy space or to favor oligarchs' preferences, but these party behaviors emerge from the joint influence of voters and donors. Similarly, special interest group members are not compelled to donate more money to a political campaign when voters are more susceptible to advertising, but theorists contend that campaign donors do behave in this way (Grossman & Helpman 1996). A political party's donation total emerges from influences like voters' reactions to advertising and oligarchs' tendency to free ride or to cooperate by making a donation.

A multi-agent-based model does not gloss over differences among agents that may affect their behavior, in order to derive an analytic equilibrium solution. Instead, it retains much of the complexity of the real world system, modeling outcomes as samples from a stochastic process in which many distinct agents evolve over time. Therefore, such an approach may be preferable to differential equation-based approaches traditionally used by political economists (Bardhan & Mookherjee, 2000). Political economics models have several characteristics that make the agent-based approach especially useful. Agents in an economic model of politics are assumed to act in self-interest. Aggregate outcomes such as parties moving toward the preferences of the median voter emerge from the interaction of many self-interested agents, as in an agent-based model. Furthermore, agents in a voter-party model have incomplete information about the state of the system, such as what each party would do if in office, or what each voter would do if the parties changed policy positions. Agent-based models are appropriate tools for modeling complex systems such as this, where there is limited information and bounded rationality. Finally, agent-based models are fully observable as they progress, so it is feasible to analyze their outcomes and changes over time in ways that may not be possible in analytic models.

Kollman, Miller, and Page (1992) presented a seminal agent-based model of interactions between voters and political parties. In this model (known as the KMP model), two parties compete for votes from a set of voters. Voters in the model know each party's position in policy space and vote for the party that is closest to them. Parties can change their positions in a discrete policy space with 15 dimensions and 7 possible values in each dimension. Parties seek either to maximize votes or to maximize a function of votes and distance from a preferred policy position. The authors tested a variety of party strategies, to evaluate which strategies would allow parties to compete most effectively and maximize expected utility over time. Laver (2005) extended this work by developing a new agent-based model of voter-party interactions. In this model, the policy space has only two dimensions, so it is easier to visualize. Laver introduced a new strategy, called the Hunter strategy, which performs well in spite of its simplicity. In this strategy, a party moves again in the same direction across policy space as in the last cycle, if its votes increased; otherwise, the party chooses a different direction at random and moves that way. In this paper, we further extend this literature by introducing Oligarchs as a new class of computational agents, and by investigating the effects of ideological preferences of the oligarchs on legislative action.

## Key Definitional Issues

Before we proceed further, it is important to clarify some definitional issues regarding the following terms: Oligarch donations, taxes and the timescale represented by each run cycle in our models.

In our work, Oligarch "donations" need not be thought of as only representing campaign contributions in an election year. Donations can also represent any effort made by a business-oriented interest group (i.e., an Oligarch) to support a Party or win its favor, at a cost to the Oligarch.

The "tax" that distributes income from Voters to Oligarchs need not be thought of as a direct subsidy to Oligarchs and tax on Voters, but as any government policy that benefits Oligarchs financially at a cost to society. Stigler (1971) lists common examples such as a protective tariff, regulatory barriers to market entry for competitors, and price controls; Smith (2000) adds increased government spending in the oligarchs' industry and corporate tax exemptions; and Olson (1965) also notes lax regulatory policies as a favor to an industry that costs society something. Hence, it is plausible that the "tax rate" in our model may fluctuate more rapidly and widely than real tax rates typically do, because it represents the effect of particularistic policies that are apt to change rapidly.

Laver (2005) explained that his model does not depict election campaigns with its discrete time steps, but rather simulates continual change in party positions and voter intentions. Kollman, Miller, and Page (1992) specified that in their model, parties learn about voter intentions of how to vote through "test elections," or opinion polls, not only through election year results. The OLIGO model, as is typical in agent-based models of democracy, lets the most recent winning Party set the policy at each step, which in the context of continuous updating can be viewed as letting the Party with greater mandate in current Voter support set policy. To validate the rate at which party support changes over time in our model, we ran our model for 120 steps after the warm-up period and compared the change in party support at each cycle to data from 10 years of monthly opinion polls on US voter support for the Democratic and Republican parties, based on data from monthly Gallup polls collected from January 2004 to December 2013 (Gallup 2014). The Gallup poll asked voters if they are Democrats, Democrat leaning, Republicans, Republican leaning, or independent. It is important to note that given our goal of creating a model that can be validated by empirical data from the Gallup poll surveys, we directly adopted and implemented these categorical definitions in our model. We added the percentages for each category, to produce a time series of voters who would vote for each party. In constructing our time series of party support in this way, we assume that surveyed voters who support or lean toward a party would be more likely to vote for a candidate of that party, if an election were held immediately.

Our results indicate that variations on our basic model called the Party-Ideo-Motivated Model and the All-Ideo-Motivated Model match the polling results closely in terms of variance over time and central tendency, as will be shown in the results section. Although 1000 steps representing 1 month each represents 83 years of real time, too long for our model to plausibly represent, this longer number of steps is helpful for steady state parameter estimation, and we will show that the behavior of our model over shorter intervals is consistent with trends in real world polling data.

# The OLIGO model

In this section we present a description of the OLIGO model, based on the ODD (Overview, Design concepts, Details) protocol (Grimm et al. 2006). The OLIGO model is implemented in NetLogo (Wilensky 1999). The model's source code can be found at http://www.openabm.org/model/3990/version/1/view. The description given here concerns our basic model, which we will refer to as the Ideo-Indifferent Model (IIM) when necessary for clarification. Two variant models, the Party-Ideo-Motivated Model (PIMM) and All-Ideo-Motivated Model (AIMM), are introduced at the end of this section.

## Purpose

To investigate how oligarchs (i.e., owners of firms in an industry with barriers to entry by new firms), could decide how much to invest in lobbying democratic parties for subsidies, and how responsive parties might be to influence from the oligarchs' campaign donations, in light of politicians' goal of increasing their votes.

## Entities, state variables, and scales

Our model contains three entity types: Oligarchs, Parties, and Voters. 5 Oligarchs, 2 Parties, and 100 Voters are created for each model run. These entity types interact as follows: Oligarchs donate money to the Party that advocates a higher redistributive tax in favor of the Oligarchs, and Oligarchs receive money from the tax enacted by the winning Party after each election. Parties receive votes from Voters and receive money in the form of campaign donations from Oligarchs. The incumbent Party sets the tax rate that redistributes money from Voters to Oligarchs. Voters give votes to the Party that appears closer to them in policy space, based on their policy position, their salience along the *olig* and *ideo* axes, and the amount of campaign funding each Party has received. Voters give money to Oligarchs based on the tax rate.

All entities in the model are positioned in a policy space, which has two axes: the *ideo* axis, representing the ideological dimension of political belief, and the *olig* axis, which represents the special interests of Oligarchs. The level of the tax on Voters that funds the Oligarchs' subsidy is determined by the incumbent Party's position on the olig policy axis. Oligarchs donate money exclusively to the Party with higher olig position, and Voters all hold the minimum position along the olig axis. Voters are distributed along the ideo axis, in contrast, from a random normal distribution with mean at the center (0), and with a standard deviation of 1/3 the radius of the space along that axis (Laver 2005). While the olig axis determines the level of redistributive tax on Voters producing a subsidy to Oligarchs, the ideo axis represents the spectrum of Voter beliefs on other political issues. As such, the ideo axis may be thought of as the "principal component" of Voter beliefs that are "orthogonal to" beliefs about the desirable level of subsidy to Oligarchs. In the real world, this axis might represent the continuum of conservative to liberal views on social issues.

Oligarchs seek to increase *profits*, which represent the difference between their subsidy and the amount spent on campaign contributions. Their only available action is to increase or decrease their campaign contribution by a fixed step size. Each Oligarch has a *gross-income* constant representing the amount of money the Oligarch earns at each time step. An Oligarch has a *donation-size* variable indicating the fraction of its gross-income it can donate to the Party that proposes a higher redistributive tax. The effect of the party policy on the net income of oligarchs is represented as *subsidy* to oligarchs. The Oligarch can change its donation-size over time to adapt to its environment. Each Oligarch has a memory of the most recent subsidy it received, and uses this knowledge to decide whether to increase or decrease its donation-size proportionally.

Note that in our model, the *gross-income* constant represents the oligarch's income *before* the effect of political policy (i.e., subsidies) and the oligarch's campaign donations. Allowing an oligarch's gross income, before subsidy and campaign donations, to vary based on factors such as incumbent party ideo position, would be an interesting extension to the model, but it is beyond the scope of the present work, given our focus on identifying the relationship between subsidies and the oligarch's donation.

| **Oligarch** |
|---|
| gross-income: Real<br>donation-size: Real<br>last-subsidy: Real<br>agent-index: Integer |

*Figure 1*. Oligarch entity variables.

Voters seek to minimize the distance of the incumbent Party from themselves along the ideo policy axis, and to decrease the tax they pay. Voters are modeled as reflex agents, which increase the salience they ascribe to the olig axis with greater probability when the tax rate is high. Voters have a position along the ideo policy axis. Voter positions along the ideo axis are initialized randomly from a normal distribution. Voters have an *olig-salience* variable (Figure 2), which represents the relative salience a Voter ascribes to the olig policy axis, compared to the ideo policy axis. For example, a Voter with an olig-salience of 1/3 will consider its distance from a Party along the olig axis to be only half as important as its distance from the Party along the ideo axis. A two-dimensional policy space with salience has been used in prior models, such as De Marchi (1999). Voters choose to vote for the Party that appears to be closer in policy space, based on the Party's policy position, the Voter's olig-salience value, and the level of campaign contributions the Party has received. The more campaign donations a Party receives from Oligarchs, the closer it will appear to each Voter, and the more likely it will be to win many votes.

| **Voter** |
|---|
| ideo: Real<br>olig-salience: Real |

*Figure 2*. Voter entity variables.

Parties seek more votes, by changing their positions in policy space. Parties have a position along the ideo and olig axes (Figure 3), and both of these positions are initialized randomly from a normal distribution with a standard deviation of 1/3 the radius of the policy space along its axes. Parties change their positions in policy space along each axis, at every time step, by a fixed step size in either direction. Parties have a memory of which way they moved at the last step, and of whether their vote totals increased or decreased, which they use to decide which way to move in the next step.

| **Party** |
|---|
| ideo: Real<br>olig: Real<br>last-increase-ideo: Boolean<br>last-increase-olig: Boolean |

*Figure 3.* Party entity variables.

We tested model variations where OLIGARCH-GDP-FRACTION was set to 0.02 or 0.1 instead of to 0.05. Results for these model variations will be shown in the results section, as increasing Oligarch income relative to Voter income seems to increase the effect of Oligarchs' influence on political outcomes. The range of Oligarch incomes used in our study is comparable to real world revenue for major US industries and corporations relative to the total economy. That is, in 2011 the total revenue of Fortune 500 companies was $10.8 trillion, while the top 12 US-based oil companies had 2013 revenues of $1.4 trillion, and Apple Inc. had 2013 revenue of $170 billion. Major oil corporation revenue was thus about

13% of the Fortune 500 total in this period, and Apple Inc. revenue was about 1.6% of the Fortune 500 total (Apple Inc. 2013; CNN Money 2013). So an OLIGARCH-GDP-FRACTION of 0.01-0.10 appears plausible as a representation of the revenue of a large corporation or industry.

Voters have a VOTER-AD-DECAY-FACTOR which affects the extent to which there is a diminishing return from increased campaign donations. VOTER-AWARENESS determines how likely a Voter is to increase its olig-salience value, when there are high tax rates. VOTER-MEMORY-STRENGTH determines how quickly a Voter's olig-salience value decreases over time, when the Voter has not noticed high tax rates recently. And VOTER-MIN-DISTANCE-SCALE is the minimum fraction of the actual distance to a Party that a Voter can be made to perceive as the true distance, when a Party has high campaign donations.

## Process overview and scheduling

After Voter, Oligarch, and Party entities have been initialized, each run of the OLIGO model consists of 1300 cycles of the step procedure, of which only the last 1000 are analyzed. We discard the first 300 cycles of each run to remove the startup effect, or initial transient bias, that is present before the run reaches a steady state; this is sometimes called the replication-deletion approach to analyzing the steady state of a system. We chose 300 cycles as the length of the discarded warm-up period through Welch's graphical approach, as described in Law and Kelton (1991). That is, we plotted the mean profit (our output variable of interest) across 100 independent runs, against the step number from 1 to 1300, and converted the plot to a moving average to smooth out noise. We then selected 300 steps as the number that seemed by inspection to precede a steady state in the moving average graph. For comparison, Laver (2005) uses 150 cycles of warm-up and 250-1000 cycles per run.

At each step, the following events happen, in order: (1) Oligarchs update their donation-size values, and thus their campaign donations; (2) Voters vote for the Party that appears closer to them in policy space; (3) the elected Party sets the tax rate, and money is redistributed from Voters to Oligarchs; (4) Voters update their olig-salience values; and (5) Parties update their positions in policy space (Figure 4). In each of these stages, entities perform their actions in a random order, but the model is structured such that the order of agents' actions is irrelevant within a given stage. One can think of the agents of any entity type as acting simultaneously. The sequence of stages in each run cycle is adapted from similar models in the literature, such as Grossman and Helpman (1996).

We use a run length of 1000 "steady state" cycles in our experiments, which may appear to be quite long. The benefit of long run duration is less noise in estimates of steady state model behavior, and this benefit may be why Laver (2005) also used runs of length up to 1000 cycles. As we discussed in an earlier section, similar to Laver (2005) and Kollman, Miller, and Page (1992), our model represents continuous adaptation of Parties and Voters between elections. Thus a single run cycle may be viewed as the shortest interval in which parties can react by adjusting their policies to a perceived change in public support. For example, an opinion polling cycle may be the closest real system equivalent to a run cycle, rather than an election cycle.

```
GO:
    UPDATE-OLIGARCH-DONATIONS
    VOTE
    REDISTRIBUTE
    UPDATE-VOTER-OLIG-SALIENCE
    UPDATE-PARTY-POLICIES
```

*Figure 4.* Main module. Pseudocode for the submodels will be introduced below.

## Design concepts: Basic principles

The OLIGO model fits into the class of “partial politico-economic models” (Frey & Schneider 1975), as it represents interactions between firms motivated by profit and political parties motivated by votes.  We model Voters and Parties with policy vectors in a two-dimensional space, where Voters vote for the Party that appears closer to them. As Blais, Nadeau, Gidengil, and Nevitte (2001) pointed out, some theorists have suggested using a “directional model” where Voters vote for the Party whose position is furthest from center in the same direction as the Voter along each policy axis, but the “proximity model” described here is more prevalent and has been more strongly supported by prior studies.

The OLIGO model follows in the tradition of Downs (1957) by modeling utility seeking voters and vote seeking parties, but with bounded rationality and uncertainty leading to heuristic behavior rather than optimizing behavior. Downs' model includes interest groups as "favor buyers" who use campaign donations or other benefits to seek influence in party policies, which reflects the Oligarchs in our model. The number of Oligarch agents in our model is fixed over time, and the Oligarchs tacitly collude, without communicating with each other, to capture legislation in their benefit (Stigler 1964). Oligarchs are thus a "privileged group" in Mancur Olson's terms, as there are few enough of them and their interests are close enough that they are all likely to invest in helping the group politically, even without being compelled to (Olson 1965). We model this special interest legislation as a direct subsidy to the Oligarchs, paid for by the Voters, who are also the taxpayers (Stigler 1971; Becker 1983; Aumann & Kurz 1977). At each step of a model run, Oligarchs decide how much to donate to the campaign of the Party that proposes a higher redistributive tax in their favor (Rouchier & Thoyer 2006; Becker 1983). Oligarchs seek to increase their expected profit, which is the difference between the subsidy they expect to receive and the donation they make to a Party (Becker 1983). We assume that Oligarchs donate only to one Party, the Party that proposes a higher redistributive tax (Brock & Magee 1978; Rubin 1975).

Oligarchs in our basic model, called the Ideo-Indifferent Model (IIM), and the Party-Ideo-Motivated Model variation (PIMM) have no preference over Party positions on the ideo axis, so the Oligarchs' donations are based entirely on Party olig positions. Recall that the winning Party olig position determines the amount of the subsidy Oligarchs receive, so only the olig axis affects Oligarchs' profits. Thus, the Oligarchs' common business interests are modeled as orthogonal to the ideological interests of Voters. We relax this assumption in the All-Ideo-Motivated Model (AIMM), a variant of the model introduced below. But there is reason to model Oligarchs' political preferences as orthogonal to the core interests of Voters on the ideo axis. Smith (2000) notes that political action committees (PACs) generally seek "particularistic" policy change, not "unifying" policy change. That is, PACs seek policies that affect a narrow constituency and are not highly salient to most voters. PACs direct their funds mainly to issues that are "narrower, less visible" and lack a "mass opinion" of the public. These policies are not partisan but can be supported equally well by a conservative or liberal party. Baumgartner, Berry, Hojnacki, Leech, and Kimball (2009) found that lobby groups' agendas were different from that of the general public, in that most issues addressed by special interest lobbyists received little press coverage and were unknown to most voters.

The political system in our model has two Parties that compete in a winner-take-all election at every step of a run, where the winning Party sets the tax rate for all Voters (Bendor, Diermeier, & Ting 2003; Brock & Magee 1978; De Marchi 1999). Building on Laver (2005), parties in our model follow a “Hunter” strategy as they seek to increase votes. In a Hunter strategy, an agent will repeat its previous action if the agent received an improved payoff in the most recent run cycle; otherwise, the agent chooses an action that is somehow the opposite of its previous one. In our model, Parties will move in policy space in the same direction as they moved in the last step, if their votes increased; otherwise, they will move in a different direction. We model Parties as being influenced indirectly by donations from a special interest lobby, as suggested by Rouchier and Thoyer (2006) and by Stigler (1971).

Voters in the OLIGO model vote for the closer Party in policy space, adjusted for the Voter’s salience value along each axis, and for the campaign donations each Party received. We use Voter salience differences along policy dimensions to model the way Voters are generally ignorant about special interest legislation unless they have been significantly harmed by that legislation, which may allow oligarchs in an industry group to capture a redistributive tax (Stigler 1971; Becker 1958; McKelvey & Ordeshook 1986; Baumgartner, Berry, Hojnacki, Leech, & Kimball 2009).

Following Stigler (1971), Voters in our model are more likely to increase the salience value they assign to the olig dimension, the higher the current tax rate. The tax assessed to a Voter can be seen as a message that has some probability of causing the Voter to increase its olig-salience, which is greater the higher the tax rate (Rubin 1975; Aidt 2000). When Voters do not receive the message that they are being taxed, their olig-salience levels decrease exponentially along a forgetting curve, similar to the exponential forgetting model pro-posed by Ebbinghaus (1913) and used recently in a political economics model by Aidt (2000).

Muis (2010) created a model where a Party's apparent distance from a Voter is reduced when the Party spends more on campaigning, up to a minimum fraction of the true distance. Our model combines all these approaches to account for the effects on Voter preferences of campaign spending and Voter salience along each policy dimension.

To simplify our model, we assume that Voters do not make campaign donations (Baron 1994). This reflects voters' status as a "large, latent group" (Olson, 1965). Olson (1965) used this phrase to indicate that voters are so numerous and diverse in their interests that they have little incentive individually to invest in political influence. We further assume that Voters do not change their policy positions over time (Muis 2010; Laver 2005); that all Voters have the same gross-income (Becker 1983); and that Voters vote in every election (Dixit & Londregan 1996).

We model Oligarchs, Voters, and Parties as myopic agents that seek to recover from a misstep that was punished in the previous cycle, or to repeat a reward that was earned in the previous cycle. The "win-stay-lose-shift" adaptive heuristic the agents follow is traditional in agent-based modeling and has been used in many other agent-based models of political behavior (Kollman, Miller, & Page 1992; Laver 2005). Even traditional non-agent-based models of political behavior assume that parties seek to maximize only present votes, not discounted future votes (Downs 1957). Laver (2005) describes his agents as "adaptive agents" that "look backward and learn from the past" using heuristics, not "hyperrational agents" that "look forward strategically." This convention in political economics models is justified by the realities of political systems, where political parties are not perfectly rational or informed. Gilens and Page (2014) note that parties lack the information and reasoning ability necessary to optimize their policies and maximize votes. Instead, parties must adapt to poll data using simple rules. Thus, it is reasonable for the OLIGO model to treat Party, Voter, and Oligarch behavior as myopic and adaptive rather than optimizing.

## Design concepts: Emergence

The OLIGO model captures the emergent behavior of Party-Voter models such as Laver (2005) and Muis (2010). Parties in these models tend to stay near the median Voter's position along the ideo policy axis, even though there is no rule specifying that they must do so.

The OLIGO model introduces a new layer of complexity by adding Oligarch agents to the system. Party positions along the olig axis do not tend to align with the median Voter's, but seem to be pulled toward the Oligarchs' position by campaign donations from the Oligarchs. Parties do not follow any rule that tells them to adjust their policy positions to seek donations. Parties seek only to increase votes, yet campaign donations indirectly lead Parties to promote Oligarchs' interests. Moreover, Oligarchs' donation levels emerge from the interac-tion among Oligarchs, Voters, and Parties, such that a change in the conditions of any agent type can alter the average level of Oligarch donations over time.

## Design concepts: Adaptation

Voters in the OLIGO model adapt over time to prevent Parties from imposing too high a redistributive tax, while also preventing Parties from drifting too far from the center along the ideo policy axis. Voters adjust the relative importance of the olig and ideo policy dimensions by increasing their olig-salience value with a certain probability, which is greater when the tax rate is high, and otherwise allowing the olig-salience value to decay exponentially along a forgetting curve.

Parties adapt to changing olig-salience values among Voters, the changing position of their opposing Party, and changing donation tendencies among Oligarchs, by updating their positions in policy space. Parties follow a Hunter rule to decide how to update their policies in pursuit of votes.

Oligarchs adapt to changing Party positions, Party chances of winning election, Voter olig-salience values, and other Oligarchs' donation sizes by updating their donation size. Oligarchs follow a Hunter strategy in pursuit of profit.

## Design concepts: Objectives

Voters vote for the Party that seems to be closest to their position along the olig and ideo axes, adjusting for the relative importance of the two dimensions. Voters assign a relative salience value to the issue dimensions based on how strongly the voter has been affected by each issue recently. Each Oligarch aims to increase net profit, which is its gross-income plus the amount it collects from a government subsidy, minus the amount it spends on lobbying. Each Party seeks to increase the votes it will receive in the next election.

## Design concepts: Learning

During a model run, all three entity types learn effective strategies for their environments. Voters learn a level of olig-salience that will protect them from high taxes while also making it likely that the incumbent Party will be near them along the ideo axis. Parties learn a policy position that will win many votes against most other positions. And Oligarchs learn a donation level that tends to increase their profit.

## Design concepts: Prediction

Agents in the OLIGO model do not attempt to predict future outcomes. They merely repeat actions that have been beneficial in the past and avoid actions that have been harmful.

## Design concepts: Sensing

Voters in the OLIGO model can detect the location of each Party in policy space, but with a distortion effect that makes a Party appear closer to the Voter, the more campaign donations it received. Voters remember how much tax they paid in the previous run step.

Parties remember how many votes they received in the previous run step, but they do not know the other Party's position, the positions of Voters, the donation policies of Oligarchs, or even their own positions. The Parties' Hunter strategy allows them to adapt to changing conditions even with limited knowledge of model state.

Oligarchs can detect which Party proposes a higher subsidy and the difference in the proposed subsidy between the two Parties. Oligarchs remember whether they increased or decreased their donation level in the previous run cycle, and whether their profit increased or decreased.

## Design concepts: Interaction

Voters in the model give votes to Parties, which influences the Parties' behavior, because the Parties attempt to increase their votes. Voters decide which Party will become the incumbent through their voting. The incumbent Party sets the tax policy, which takes money from Voters and redistributes it to Oligarchs. Oligarchs donate money to the Party that proposes a higher tax, which influences Voters' vote choices and indirectly affects the policy positions of Parties.

## Design concepts: Stochasticity

Many elements of the OLIGO model are randomized, or stochastic. In the initialization of a model run, Party positions are picked from a random normal distribution along the ideo and olig axes, and Voter ideo positions are picked from a random normal distribution, as in Laver (2005). Oligarch gross-incomes are drawn from a random lognormal distribution, a common model of income distribution (Salem & Mount 1974).

During a run cycle, Parties and Oligarchs act as Hunter agents, which means that if their previous action did not improve their outcome, they will choose a different action at random. Voters act as probabilistic reflex agents when they update their olig-salience values, as they will not always "notice" a high tax rate and update their olig-salience in response, although they likelihood a Voter will increase its olig-salience strictly increases with the tax rate. In the case of a tie in an election or a voter equidistant from both Parties, the winning Party or vote recipient is decided by a fair random draw, as in Bendor, Diermeier, and Ting (2003).

## Design concepts: Observation

For each cycle of each model run, we recorded the Parties' mean position along each policy axis, the difference between the Parties' positions along each policy axis, the tax rate, the Voters' mean olig-salience value, the Oligarchs' mean do-nation size, whether the incumbent Party was closer to the center of the ideo axis than its opponent, and whether the incumbent Party had a higher olig value that its opponent.

## Initialization

Before a run begins, 2 Parties, 100 Voters, and 5 Oligarchs are created. We tested the model with 50 Voters and 5 Oligarchs, as well as with 100 Voters and 10 or 50 Oligarchs, while total Voter income and total Oligarch income were held constant. These variations did not produce meaningful differences in outcomes, so the model appears robust to changes in the ratio of Voter count to Oligarch count.

As noted above, Parties are initialized with ideo and olig positions drawn from a normal distribution with mean 0 and standard deviation 1/3 the radius of each policy axis, as in Laver (2005). Voter ideo positions are drawn in the same manner. A redraw is taken for any value outside the range from MIN-POS (-100) to MAX-POS (100). Voters' olig-salience is initialized to 0.

We also tested a version of the model where Voter ideo positions were drawn from a bimodal distribution, but the results were not significantly different from results with the unimodal distribution. We chose a bimodal distribution as an alternative because it can represent a polarized population of voters, as suggested by Downs (1957). We constructed the distribution using a mixture of Gaussians to represent polarized voter positions as in (DiMaggio, Evans & Bryson 1996). In our bimodal distribution of Voter ideo positions, we constructed an equal mixture of two Gaussian distributions to have the same overall standard deviation as our unimodal distribution (33.3), where each of the two Gaussians had a standard deviation of half that amount (16.7). We arrived at a Gaussian mixture with norms 28.9 and -28.9, standard deviations of 16.7, and mixing parameter 0.5.

Oligarchs are initialized such that their gross-income values are drawn from a lognormal distribution, which is commonly used to model the distribution of incomes (Salem & Mount 1974). A lognormal distribution is appropriate for modeling income distributions because unlike a normal distribution, it assigns probability zero to negative values and near zero probability to very small positive values; and it has been shown empirically to represent actual income distributions accurately. Although only 5 Oligarchs are present in the model whose results are shown, we tested the model with 10 Oligarchs and 50 Oligarchs to ensure robustness with the small number of samples from a lognormal distribution. There was no meaningful difference in results with greater numbers of Oligarchs, while the total Oligarch income was held constant. Oligarch counts in the range of 5 to 50 appear reasonable as a model of real privileged groups, as Olson (1965) notes that most trade associations have 25-50 members, with 1/3 having fewer than 20, and most contributions to a trade association come from a small subset of the members.

## Input data

The model does not use data from external sources.

## Model Variant: Party-Ideo-Motivated Model

We present here the Party-Ideo-Motivated Model (PIMM), the first of two variants on the base OLIGO model that was presented above. The PIMM was developed to remove a couple of simplifying assumptions of the Ideo-Indifferent Model (IIM) and to improve the model's fit for real world data of changes in party support over time.

The PIMM introduces an ideological motivation for each Party, in addition to the Parties' vote seeking motives. The two Parties are distinguished as the blue, liberal Party and the red, conservative Party. The blue Party has a bliss point along the ideo axis of -33, the red Party at +33. Moreover, the Parties are unwilling to adopt ideo policy positions on the "other side" of the opposing Party's position. For example, the red Party will not adopt an ideo position less than or equal to the blue Party's.

The PIMM makes some Voters "loyalists", instead of having all Voters be "swing" Voters as in the IIM. Loyalist Voters are not influenced by campaign spending or the olig position of Parties, but vote based entirely on the positions of the Parties along the ideo axis. Theorists including Downs (1957) note that many voters act as loyalists, in that they habitually vote for the same party or the party that matches their ideology best.

In the PIMM, a SWING-VOTER-FRACTION parameter of 0.33 indicates that 2/3 of Voters will be loyalists, and only 1/3 of Voters will alter their votes based on campaign donations or Party olig position. We chose these parameter settings based on a Pew Research study that shows approximately 33% of registered voters in the United States were swing voters, as of 2008. Since 2008, the fraction of voters classified as swing voters has become smaller, so this parameter setting is conservative. (Pew Research 2012.)

Initial positions of the two Parties along the ideo axis in the PIMM are at the Parties' bliss points of -33 and +33, rather than being drawn from a 0-mean Gaussian distribution.

In the PIMM, if a Party's position update function directs it to adopt an ideo position that is on the "wrong side" of its opposing Party's current policy, the Party will instead keep its ideo position unchanged in that cycle. If a Party's ideo position is on the "wrong side" of its bliss point (lower than +33 or higher than -33), the Party adds a bias term to its updated ideo position, which depends on the distance from its bliss point. The bias term is 0 at the bliss point and approaches PARTY-EPSILON at the edge of policy space, so that in the worst case this bias completely cancels a move in the "wrong direction." For the red Party, the bias term is calculated as PARTY-EPSILON / (1 + e ^ ((50 - (33 - ideo)) / 8)). The blue Party function is similar.

## Model Variant: All-Ideo-Motivated Model

Our last model variant is the All-Ideo-Motivated Model (AIMM), which is derived from the Party-Ideo-Motivated Model (PIMM), but assigns an ideo preference in favor of "conservative" policies to the Oligarchs. It is reasonable to assign Oligarchs a preference for conservative policies because there is real world evidence that business interest groups typically hold moderately conservative views on "unifying" issues (Smith 2000). Smith shows that business interest groups usually adopt conservative stances on policies that have broad implications for all market sectors. But Smith notes that a business-oriented interest group may oppose a conservative, free market policy if the policy could hurt its profits. For example, an industry lobbying group might support increased government spending, subsidies, or regulatory barriers to entry for the group's own industry, even though such policies tend to increase taxes and government and spending.

Note that in the AIMM, Oligarchs update their donation-level values just as in the other models, based on changes in profits. Oligarchs in the AIMM do not account for the ideo axis position of the incumbent Party when deciding how to adjust donation levels, only in deciding which Party to donate to. Oligarchs in the AIMM can be thought of as industry leaders who seek to maximize profits, but believe that one way to do so is by supporting the Party with a certain ideology, regardless of the Party's current policy on industry-related issues. Thus, the Oligarchs are more likely to donate to the Party with a preferred ideo axis position, but if campaign donations do not lead to increased profits, the donation level will be reduced.

In our implementation of the AIMM, Oligarchs have a preferred ideo position of OLIGARCH-IDEO = 33, which is moderately conservative. Oligarchs are assigned an OLIGARCH-IDEO-OLIG-RATIO = 0.5, which controls the relative importance of olig and ideo position for Oligarchs considering which Party to support. Because the two axes are weighted equally in our parameterization, Oligarchs donate to the Party whose distance from [olig=100, ideo=33] is least by Manhattan distance. The amount of the donation is based on a function of the Parties' distance from the Oligarchs' bliss point as in IIM and PIMM, but in the AIMM a Party's position along both axes is considered together, leading to a more complex calculation. In brief, the PARTY-OLIG-DIFFERENCE is replaced by PARTY-OVERALL-DIFFERENCE. PARTY-OVERALL-DIFFERENCE is also in the range [0, 1], but it incorporates both policy axes by taking the absolute value of the mean of red's distance minus blue's distance, over each the two axes. To ensure the result is in [0, 1], the difference in distances between red and blue along each axis is divided by the maximum possible difference along that axis before the mean is taken.

# Simulation experiments

In this section we present: (1) the outcome variables used to describe the results of model runs, (2) an example run of the model that should help readers to understand how the model works, (3) the verification tests we performed to ensure that the model is properly calibrated, and (4) the results of hypothesis tests where we attempted to confirm that the model conforms to prior theory on the emergence of Oligarch and Voter behavior.

All statistical analyses were performed using the R statistics program. Experiments were carried out in NetLogo using the BehaviorSpace extension.

## Outcome variables

We used three statistics to measure Oligarchs' success in capturing a subsidy from the government: the mean voter-tax-rate, the mean olig position of the Parties, and the fraction of run cycles in which the winning Party had both the greater olig value and the less central ideo value (in other words, the greater ideo position absolute value). The greater any of these statistics, the more successful Oligarchs were at influencing election outcomes.

To measure the efficiency of Oligarchs at capturing a subsidy, we used the mean profit earned by Oligarchs, which is calculated as the total tax collected, minus the total campaign donations by Oligarchs, all divided by the number of Oligarchs.

To measure the campaign donations of Oligarchs, we used the mean fraction of gross-income donated by Oligarchs, which is the mean of the Oligarchs' donation-size parameter, times the PARTY-OLIG-DIFFERENCE-SCALED factor.

To measure how successful Voters were at keeping Parties near the center of policy space in the ideo dimension, we used the absolute value of the mean of the Parties' ideo values.

## Typical run description

We present results from a typical run of the OLIGO model, to help readers understand basic model behavior. In the example run, the model was run once for 1300 cycles. In Figures 5, 6, and 9, we present time series data on the positions of the two Parties, the mean fraction of gross-income donated to campaigns by the Oligarchs, and the Oligarchs' mean profit.

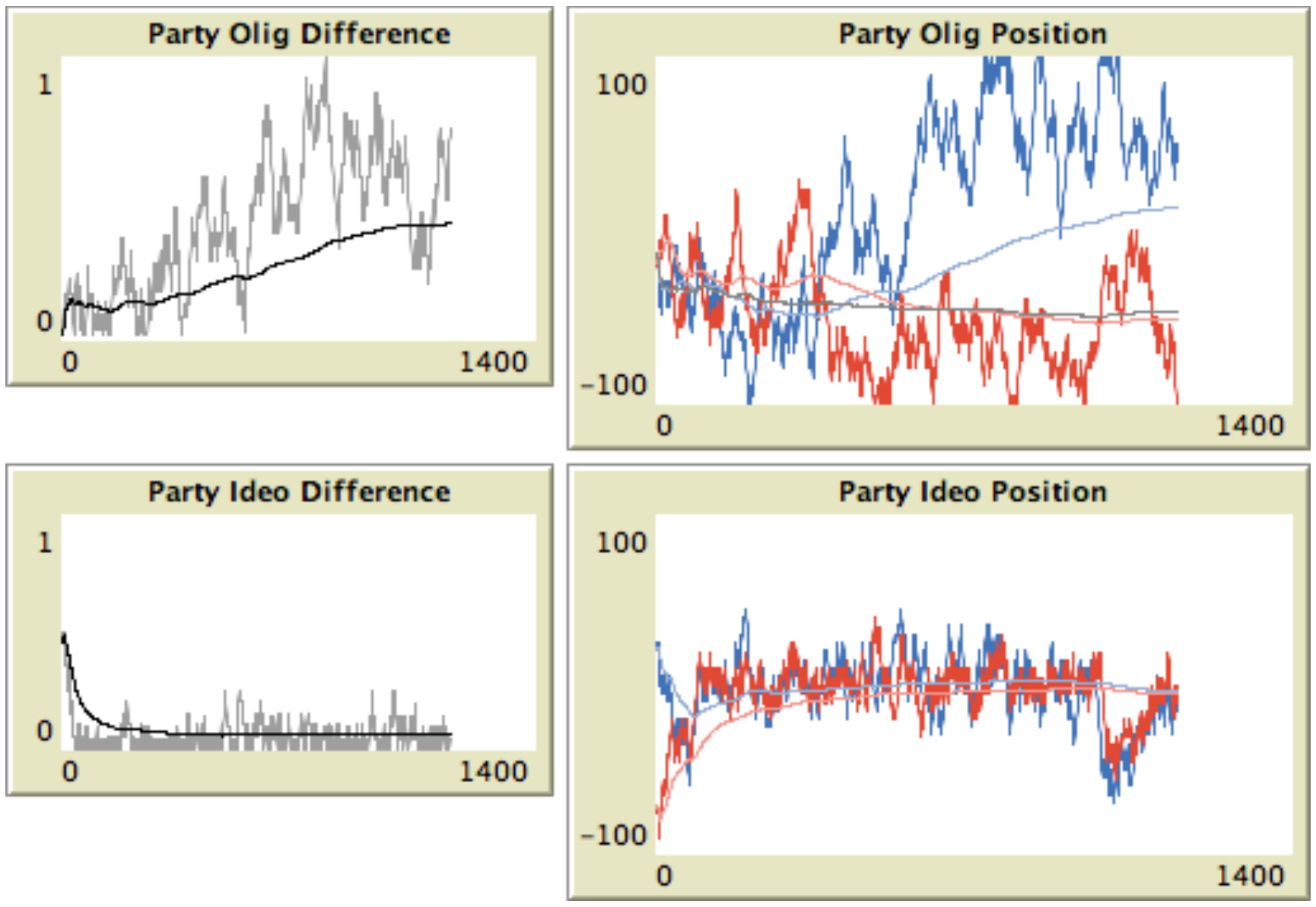

*Figure 5*. Party position versus time in a typical model run. Left: Gray lines show current difference in olig (top) or ideo (bottom) position between Parties. Black lines show all-time averages.
Right: Dark red and blue lines show Parties' current positions in olig (top) or ideo (bottom) dimension. Light red and blue lines show all-time averages for each Party. Gray lines show all-time averages for the incumbent Party.

Note that in a typical model run, Party ideo values tend to remain near the central value of zero, and they tend to remain close together. This makes sense, because by the median voter theorem, Voters evenly distributed about zero on a policy axis will tend to vote for the Party whose position is closest to zero. A Party that is much further than its opponent from the center will tend to be punished by Voters, many of which will find the other Party closer in policy space and vote for that Party. If a Party is far from the center along the ideo axis, it will be relatively easy for its opponent to win more votes, because most positions in policy space will be preferred by the median Voter to that Party's position.

Parties' olig values tend to be farther apart than their ideo values. This is because Voters typically have low olig-salience values, so they will not punish a Party as heavily for being farther away along the olig axis as for being farther away along the ideo axis. This means that a Party could receive more marginal benefit from the greater donations it receives from Oligarchs by taking a higher ideo position, than the marginal cost of taking that less popular position with Voters. As a result, Parties are able to adopt olig values greater than the minimal value, even though Voters prefer Parties that adopt the minimum.

Often during an OLIGO model run, one Party occupies a significantly higher olig axis position than the other. Although both Parties' mean olig axis positions have a 95% confidence interval that includes 0, there is a mean difference between the two Parties' olig positions in a particular step of 56.4, with 95% confidence interval [52.5, 60.4]. Recall that the full range of olig space is [-100, 100].

Thus, the two Parties tend to propose significantly different tax rates (determined by olig axis position) in a given time step, although the Parties may later switch orientations along this axis so that neither promotes higher taxes than the other on average. Individual model runs, as shown in Figure 6, tend to show occasional spikes in tax rate where a Party with extremely high olig value is elected, followed by a spike in Voter olig-salience, and a reduction in tax rates as the elected Party moves to a lower olig value or is voted out again.

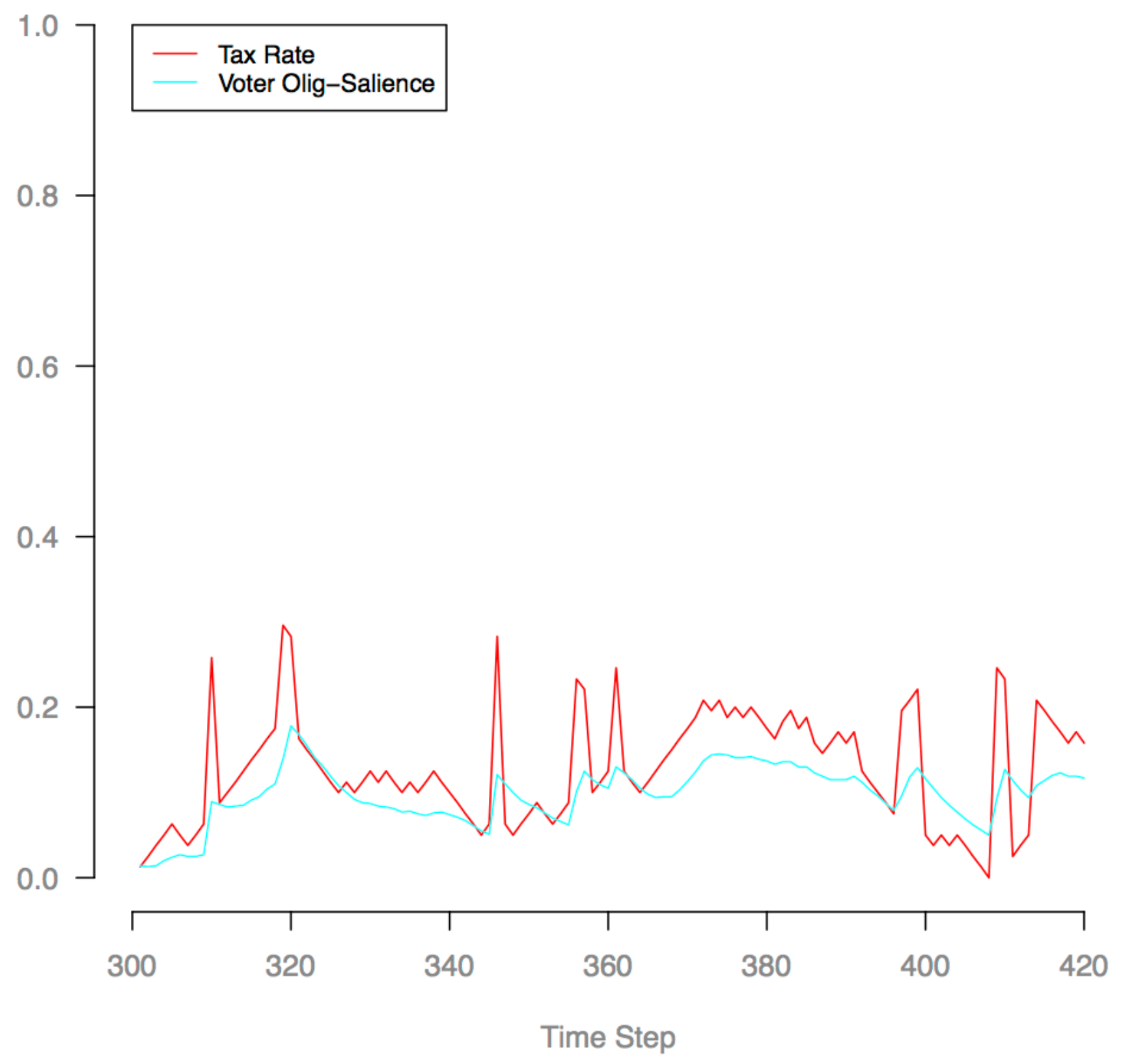

*Figure 6*. Mean Voter olig-salience and tax rate, versus time step, over a single run of 120 steps after the 300-step warm-up period.

One Party typically breaks away from the other along the olig axis, adopting a much greater olig value than its opponent, likely because only one Party can receive donations from the Oligarchs at a time, so there is more marginal benefit to this Party in seeking greater donations for itself, than to its opponent in increasing its olig position to reduce the Oligarch donation amount. In contrast, Parties frequently trade places being closer to the center along the ideo dimension. For a video of a typical model run, see Figure 7. A video of the Party-Ideo-Motivated model appears for comparison, as Figure 8.

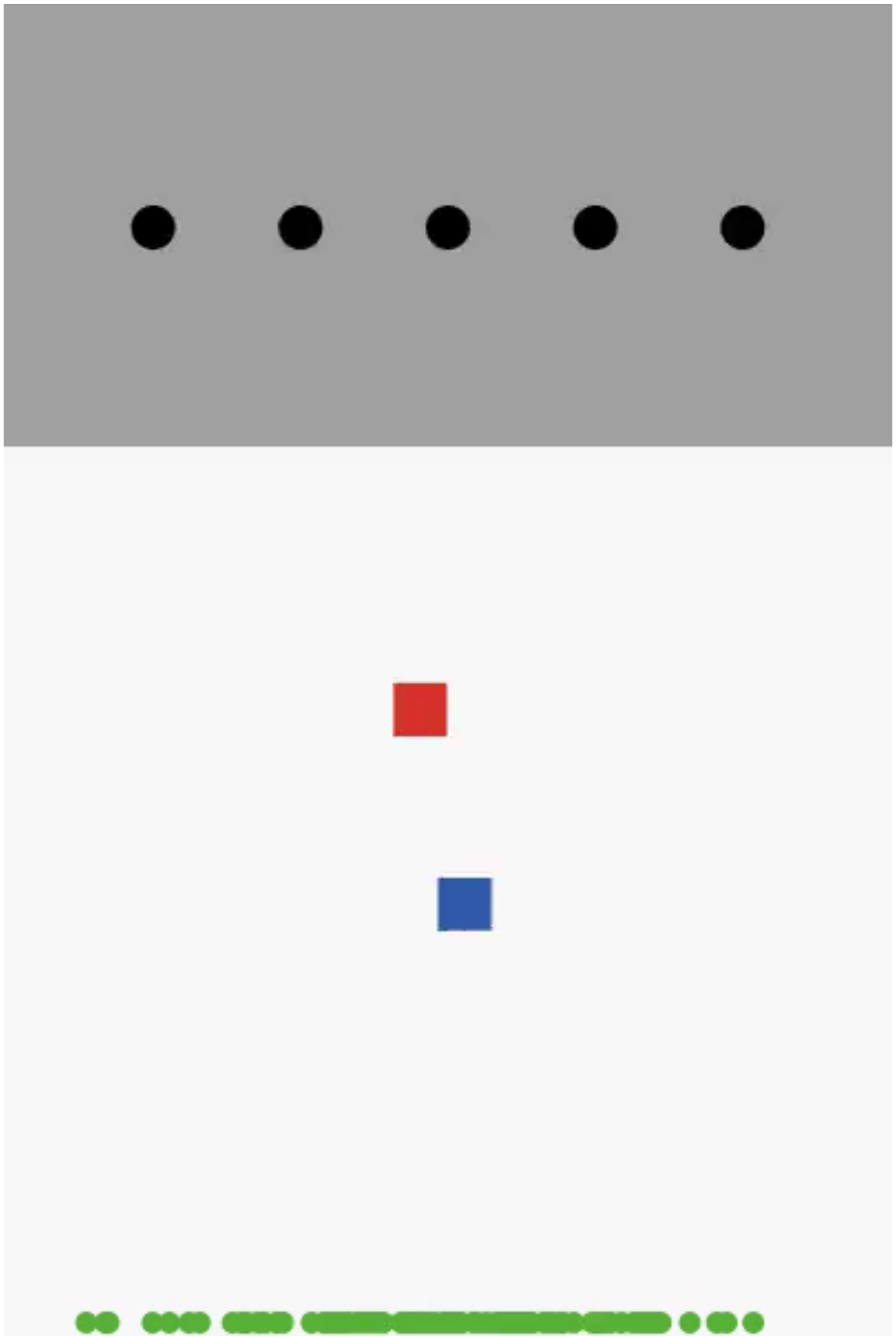

*Figure 7*. Video of a base model (IIM) run of 500 steps. The five top circles are Oligarchs. Oligarch color shows which Party was recently supported, radius is proportional to amount donated, and label shows donation-size. The two squares are Parties. Party width is proportional to votes received, label indicates if votes increased or decreased, and outline shows the incumbent. The 100 bottom circles are Voters. Voter color shows which Party was voted for. The ideo axis is vertical and the olig axis is horizontal. Each Party and Voter's location corresponds to its position in policy space.

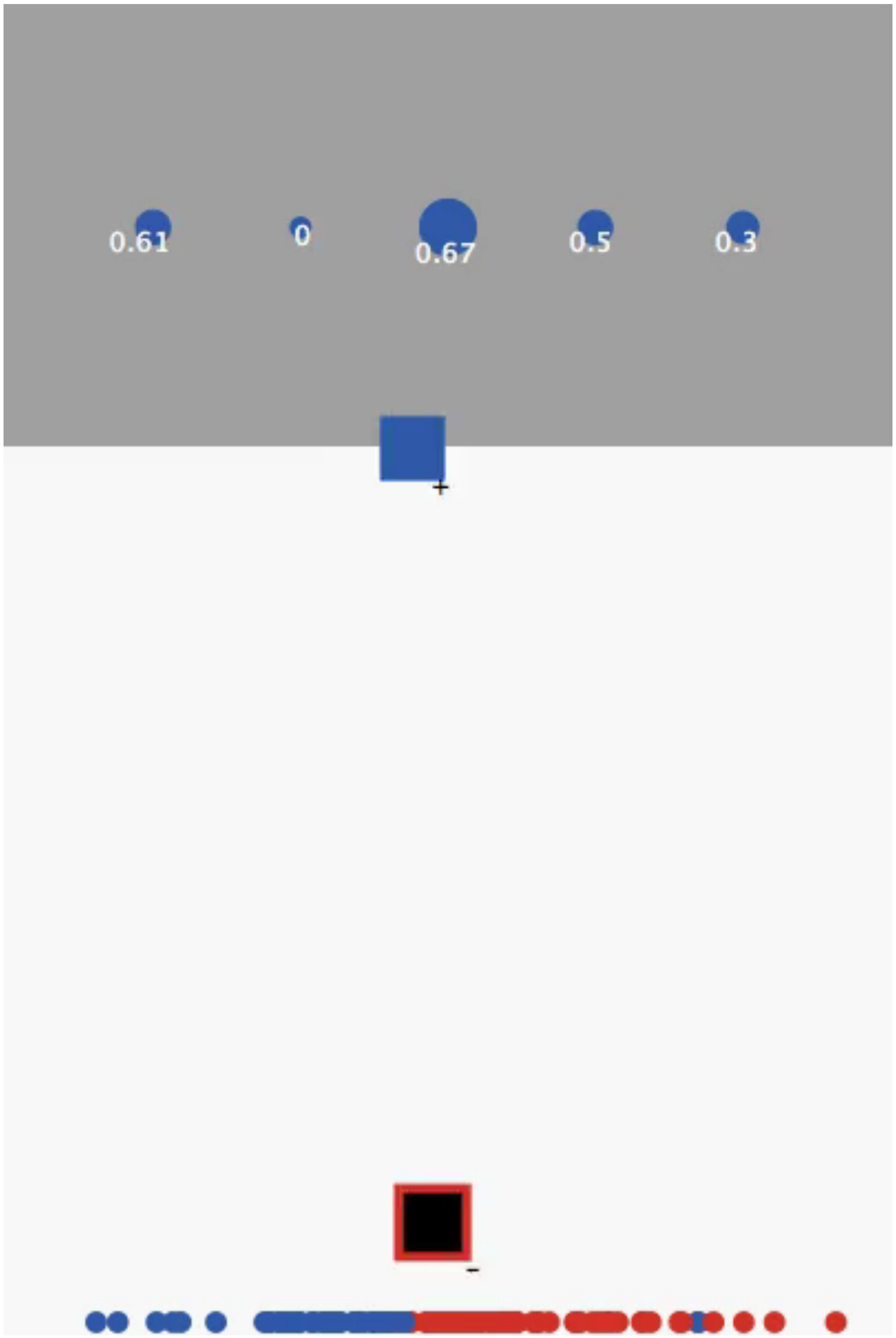

*Figure 8*. Video of a Party-Ideo-Motivated Model run, beginning after the 300-step warm-up period.

Time series plots averaged over 100 runs of the OLIGO model suggest that a 300 cycle warm-up period was sufficient for most model statistics to become stable, in the sense that the means of most model statistics across runs were not dependent on the time step. In particular, the mean olig and ideo positions of the Parties, the tax rate, and the mean olig-salience of Voters had nearly the same means across runs, regardless of the time step after warm-up (Figure 9). For mean Oligarch profit, however, there was change over time; this variable did not seem to attain a stable value over time steps, even in the mean across runs (Figure 10), although its drift appeared to stop after 300 steps.

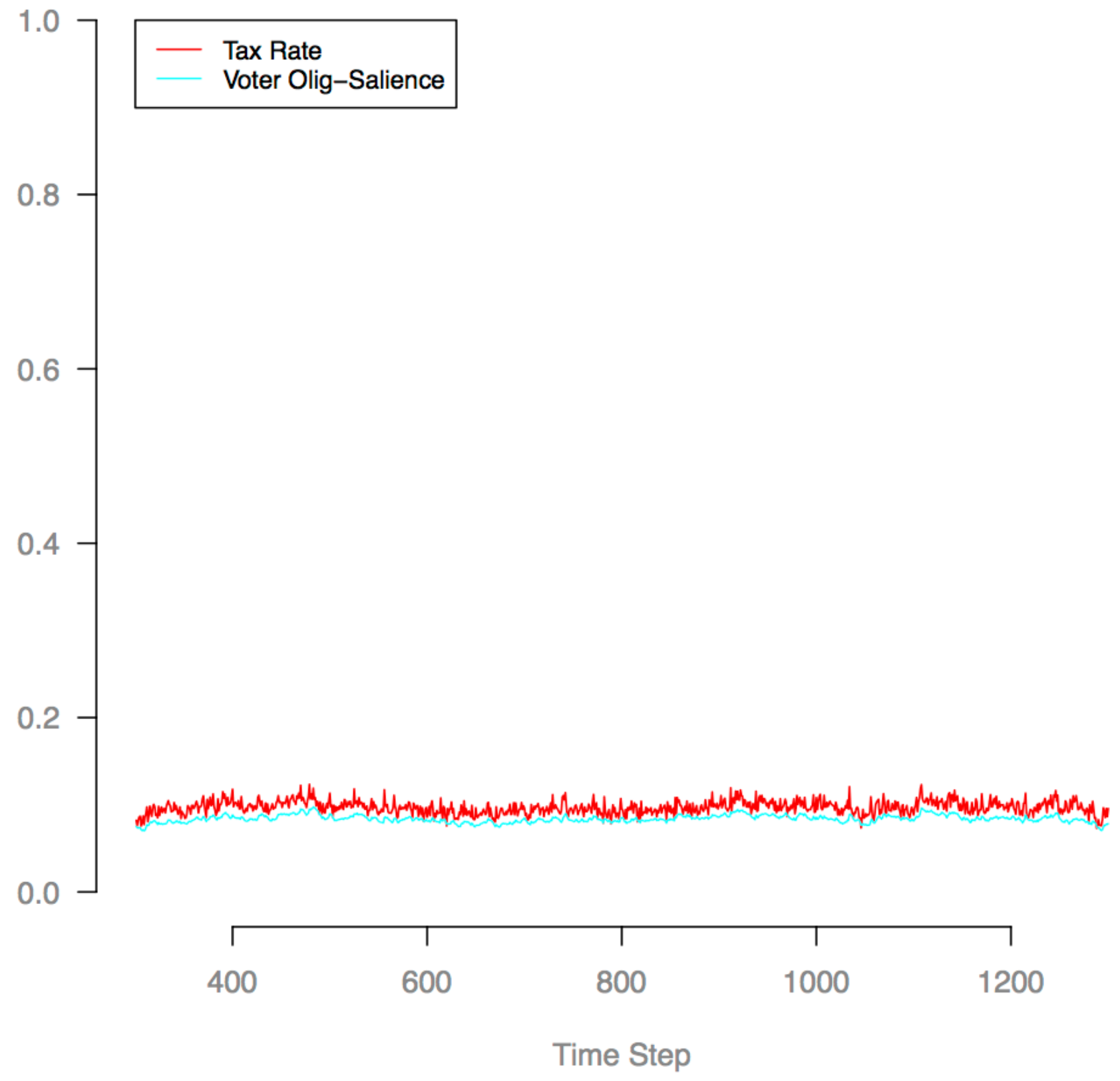

*Figure 9*. Mean Voter olig-salience and tax rate, versus time step, averaged over 100 runs.

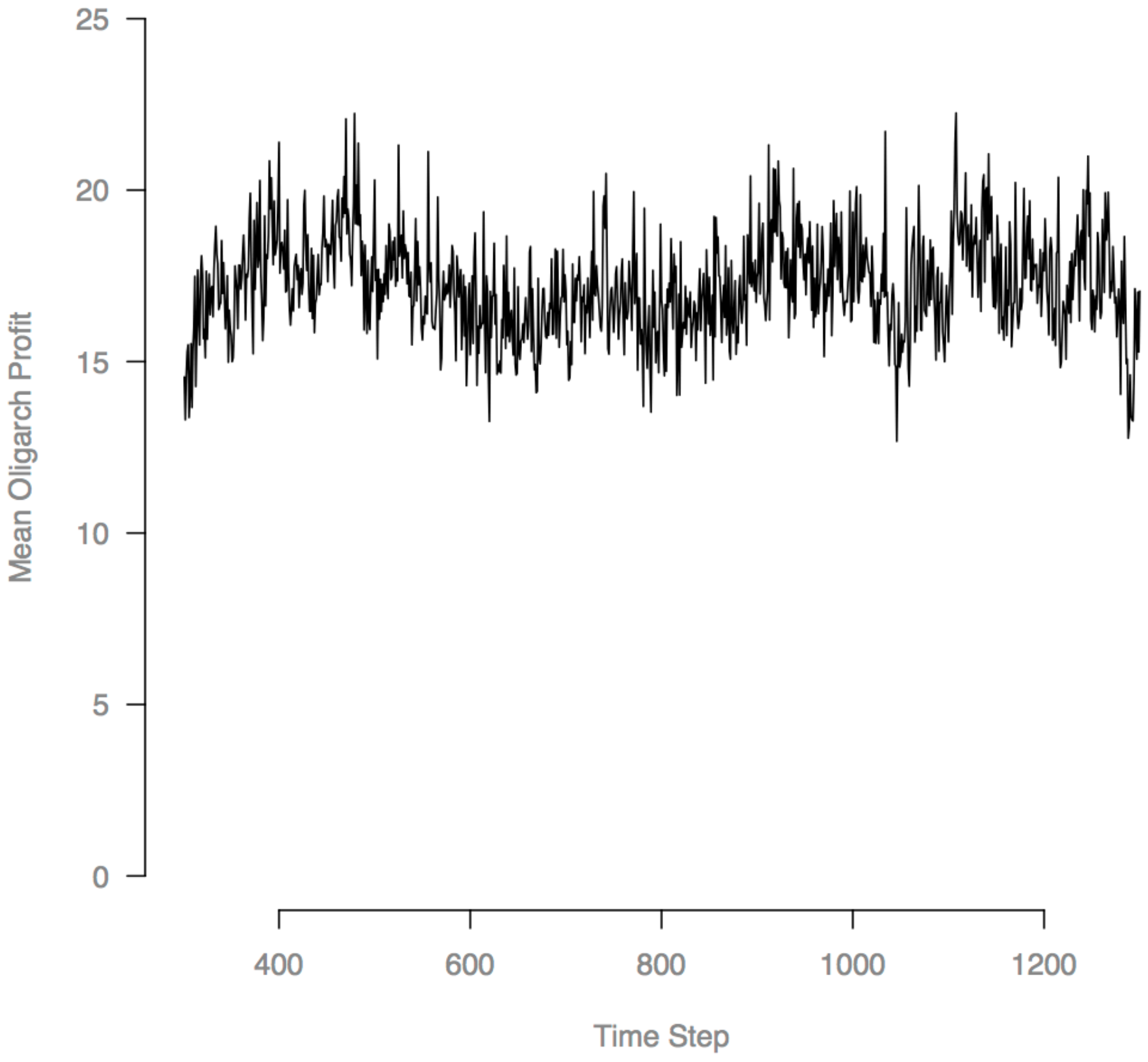

*Figure 10*. Mean Oligarch profit versus time step, averaged over 100 runs.

## Experimental methods

We performed six computational experiments with the IIM version of the OLIGO model (Figure 11).

1. We ran the base model 100 times to obtain summary statistics.
2. To explore how Oligarchs and Voters would act if their donations or olig-salience had no effect on Party olig positions, we modified our model to set Party olig positions at fixed values that could not change during a run; we ran this model 100 times.
3. To study the effect of fixing Oligarch donation-size at various levels, we ran a parameter sweep for the initial Oligarch donation-size, and we fixed Oligarch donation-size in place so that Oligarchs could not change their donation-size from the assigned value. We ran the model 20 times at each value.

4. We ran a parameter sweep for VOTER-AD-DECAY-FACTOR, which affects how strongly Voters react to campaign donations. We ran the model 20 times at each value.
5. We ran a parameter sweep over the initial olig-salience of Voters, in a model variation where Voters could not change their olig-salience from its original value. We ran the model 20 times at each value.
6. We ran a parameter sweep over the VOTER-MEMORY-STRENGTH, which affects how long Voters maintain a high olig-salience after increasing their olig-salience due to high taxes. We ran the model 20 times at each value.

| |
|---|
| **Base model, such as Ideo-Indifferent Model (IIM)**<br>1300 cycles per run (discard first 300)<br>100 runs |
| **Fixed Party olig**<br>1300 cycles per run (discard first 300)<br>100 runs |
| **Fixed Oligarch donation-size**<br>**OLIGARCH-INITIAL-DONATION parameter sweep**<br>OLIGARCH-INITIAL-DONATION values: $(0, 0.2, 0.4, 0.6, 0.8, 1)$<br>1300 cycles per run (discard first 300)<br>20 runs per parameter value, 120 runs total |
| **VOTER-AD-DECAY-FACTOR parameter sweep**<br>VOTER-AD-DECAY-FACTOR values: $(-0.12, -0.1, -0.08, -0.06, -0.04, -0.02)$<br>1300 cycles per run (discard first 300)<br>20 runs per parameter value, 120 runs total |
| **Fixed Voter olig-salience**<br>**VOTER-INITIAL-SALIENCE parameter sweep**<br>VOTER-INITIAL-SALIENCE values: $(0, 0.2, 0.4, 0.6, 0.8, 1)$<br>1300 cycles per run (discard first 300)<br>20 runs per parameter value, 120 runs total |
| **VOTER-MEMORY-STRENGTH parameter sweep**<br>VOTER-MEMORY-STRENGTH values: $(0.1, 0.3, 0.5, 0.7, 0.9)$<br>1300 cycles per run (discard first 300)<br>20 runs per parameter value, 100 runs total |

*Figure 11*. Experimental conditions.

We also analyzed two variants of the IIM, which we call the Party-Ideo-Motivated Model (PIMM) and All-Ideo-Motivated Model (AIMM). Recall that in the Party-Ideo-Motivated Model, unlike in the IIM, each Party has a preferred ideo axis position, and some Voters are loyalists instead of swing voters. In the All-Ideo-Motivated Model, changes from the PIMM are retained, and additionally, Oligarchs have a preferred ideo position that favors the red, conservative Party. Oligarchs in the AIMM will donate to the red Party even if the blue Party's olig position is greater, as long as this difference in olig position is outweighed by the difference in ideo position.

To evaluate the results of our experiments, we used hypothesis tests that make minimal assumptions about the distribution of results. We used Welch two-sample t-tests to test for different means between samples, where it seemed likely that the samples had different variances, so a conventional two-sample t-test would not work properly. We used cross-correlation, with lags of -5 to 5 run steps, to look for effects of one variable on another in our default model run. We chose a lagged approach so that we could detect relationships between variables that require multiple run steps to emerge. To evaluate the effect of the sweep parameter on other variables in our parameter sweeps, we used Spearman correlation to look for a monotonic relationship between the variables that might be nonlinear.

## Model verification

We performed a series of verification tests to ensure that the OLIGO model behaved in accordance with prior results from political economics as well as with common sense.

According to the median voter theorem, Voters with normally distributed preferences will tend to elect a Party whose position in policy space is aligned with the preferences of the median Voter. Because the OLIGO model initializes Voters with ideo values drawn from a random normal distribution centered on zero, the mean ideo value of Parties should be zero when averaged over many runs, because Parties seek to increase the votes they receive. A two-tailed Welch's t-test on the 100 default runs could not reject the hypothesis that the mean Party ideo position was equal to zero, $p = 0.84$, 95% confidence interval (-1.042, 0.842) in a possible range of ideo values from -100 to 100.

If the base model is properly calibrated and scaled, Oligarch donations should not dominate the effect of Voter preferences. In other words, Oligarchs should not be able to capture severe redistributive taxes. Specifically, we expected the mean Party olig to be below zero (i.e., closer to Voters' preferred value than to Oligarchs') and the mean Party olig to be below the mean Party ideo value. A one-tailed Welch's t-test confirmed that the mean Party olig was indeed less than zero, $p < 2.2e\text{-}16$, $M = -45.3$.

For our model to perform reliably, it needs to be non-degenerate – that is, all agents must take some action at equilibrium that has an effect on model outcomes. To verify the non-degeneracy of the base model, we used one-tailed Welch's t-tests to verify all of the following conditions, $p < 2.2e\text{-}16$:

- mean olig position of Parties > MIN-POS
- mean olig-salience of Voters > 0
- mean tax rate < MAX-TAX
- mean Oligarch donation-size > 0

## Results: Base model experiment

Cross-correlations revealed that in runs of the base OLIGO model, called the Ideo-Indifferent Model or IIM for clarity, both tax rate and mean Oligarch profit were positively correlated with mean olig-salience of Voters at small lags (Figure 12). A two-sided 95% confidence interval for the cross-correlation at each lag was computed by treating the cross-correlations at that lag from different runs as IID samples from a random variable and computing the t confidence interval, based on 100 runs. The correlation had a maximum over lags at 0.695 in [0.677, 0.713] for tax rate to olig-salience, and at 0.678 in [0.659, 0.698] for profit to olig-salience; both of these correlation sizes are “strong.” Tax rate led changes in olig-salience, with much larger correlations at small negative lags than at positive lags. Profit also led changes in olig-salience. These results make sense, because the model is specified such that Voters are more likely to “notice” the redistributive tax and increase their olig-salience when the tax rate is high. When tax rates are high, profits tend to be high also, because profits are computed by subtracting donations from taxes collected.

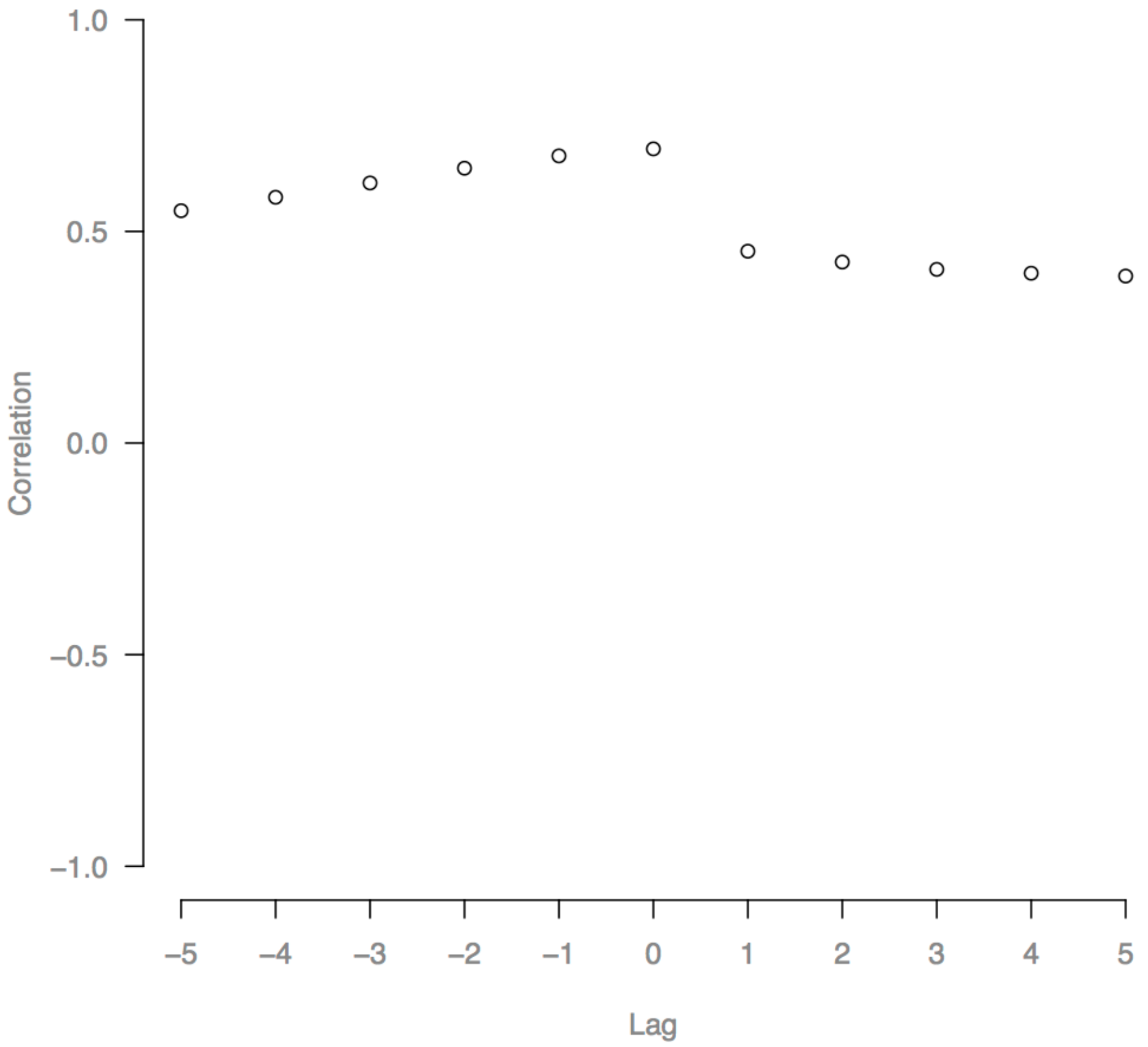

*Figure 12*. Cross-correlation of tax rate to mean Voter olig-salience, in the base model. Cross-correlations were taken for each run and averaged over 100 runs.

The mean donation-size of Oligarchs was positively correlated with mean Voter olig-salience, but the maximum correlation over lags was only 0.243 in [0.201, 0.285], which is “weak” (Figure 13). Neither variable of donation-size and olig-salience appeared to lead the other, as the cross-correlations were nearly constant over small lags. This indicates that although Voter olig-salience tended to be high when Oligarch donations were high, there was little immediate effect of change in one variable on the other, for the small changes that typically occurred in our base model. Indeed, our model does not include rules that directly link these two variables, so it is understandable that they did not produce a strongly peaked cross-correlation plot.

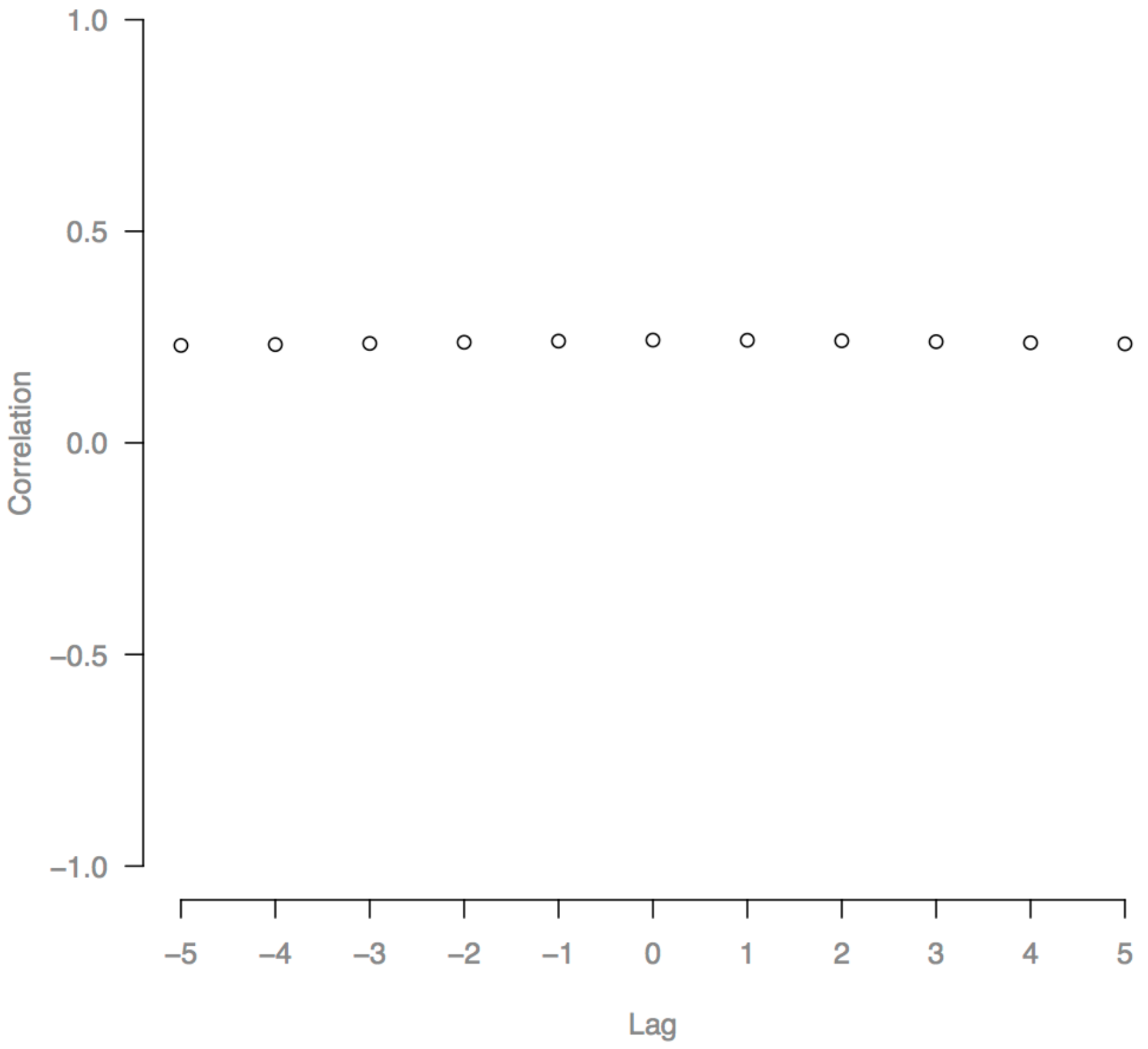

*Figure 13*. Cross-correlation of mean Voter olig-salience to mean Oligarch donation-size, in the base model. Cross-correlations were taken for each run and averaged over 100 runs.

To evaluate the robustness of the results from IIM over changes to input parameters, we performed the same statistical analyses for several variations of IIM. We modified IIM in one of several ways: (1) IIM with 10 Oligarchs instead of 5; (2) IIM with 50 Oligarchs instead of 5; (3) IIM with 50 Voters instead of 100; (4) IIM with a bimodal distribution of Voter ideo positions (mixture of 2 Gaussians) instead of a normal distribution; and (5) IIM where Oligarchs are free to keep their donation level the same from one step to the next instead of increasing or decreasing it. In all of these variations, output parameter results were not significantly different from in the base IIM. See Table 1. It does appear that introducing the null action for Oligarchs leads to slightly increased Oligarch profits and taxes, with increased donation rates. But other outcome measures are very similar between the null option model and the base IIM, so these differences do not appear to alter our key findings. Note that when Oligarch and Voter counts were changed, the total income of each group was kept the same.

| Model | Max red vote | Mean abs red vote change | Mean profit | Mean donation | Mean tax | Mean salience |
|---|---|---|---|---|---|---|
| IIM | 93.35 | 7.81 | 17.20 | 0.11 | 0.10 | 0.08 |
| 50 Voters | 95.12* | 7.92* | 17.13 | 0.10 | 0.10 | 0.08 |
| 10 Oligarchs | 93.39 | 8.03 | 16.78** | 0.10 | 0.09 | 0.08 |
| 50 Oligarchs | 92.46 | 7.63 | 16.2*** | 0.09 | 0.09 | 0.08 |
| Bimodal voters | 93.58 | 6.74 | 17.04 | 0.10 | 0.10 | 0.08 |
| Null option | 89.18 | 7.59 | 19.84 | 0.22 | 0.12 | 0.11 |

*Table 1*. Comparison of Ideo-Indifferent Model, and variations with 50 Voters, 10 Oligarchs, 50 Oligarchs, bimodal distribution of Voter ideo position, and null donation change option for Oligarchs.
* 50-Voter max and mean votes were multiplied by 100/50 for comparison.
** 10-Oligarch mean profits were multiplied by 10/5 for comparison.
*** 50-Oligarch mean profits were multiplied by 50/5 for comparison.

Our model does appear sensitive to a change in the fraction of total income assigned to the Oligarchs (Table 2). We performed a sensitivity analysis in which the IIM was analyzed with Oligarchs receiving 2% or 10% of total income, instead of 5% as in the base version of IIM. We expected that increasing Oligarch income would allow Oligarchs to donate more money to Party campaigns, leading to greater advertising effects, higher mean Party olig values, and greater Oligarch profits. This hypothesis appears to be supported by the results, based on a comparison of outcomes at 2%, 5%, and 10% Oligarch income fractions. Increased Oligarch income increased mean taxes by 0.018 from 2% to 5% (Cohen's d = 1.05) and by 0.037 from 5% to 10% (d = 1.36). Mean Party olig positions increased by 10.8 from 2% to 5% (d = 0.793), and by 11.1 from 5% to 10% (d = 0.77). Mean Oligarch profit increased by 2.17 from 2% to 5% (d = 0.672), and by 3.82 from 5% to 10% (d = 0.836). All of these Cohen's d values are "large."

| Model | Mean profit | Mean donation | Mean tax | Mean salience |
|---|---|---|---|---|
| IIM | 17.20 | 0.11 | 0.10 | 0.08 |
| PIMM | 17.60 | 0.10 | 0.10 | 0.08 |
| AIMM | 15.55 | 0.06 | 0.09 | 0.07 |
| IIM fixed Party olig | 41.0 | 0.01 | . | 0.17 |
| IIM 2% Oligarch income | 15.04 | 0.08 | 0.08 | 0.06 |
| IIM 10% Oligarch income | 21.03 | 0.14 | 0.13 | 0.12 |

*Table 2.* Comparison of Ideo-Indifferent Model, Party-Ideo-Motivated Model, All-Ideo-Motivated Model, IIM with fixed Party olig position, IIM with 2% of income to Oligarchs, and IIM with 10% of income to Oligarchs. Mean donation is the mean donation level parameter over all Oligarchs. Mean tax was not recorded for the fixed Party olig experiment.

## Results: Fixed Party olig experiment

In the fixed Party olig experiment, Parties were initialized with randomly drawn olig and ideo positions as usual, but their olig values were permanently fixed to their initial values. As a result, Voters could not induce Parties to reduce their olig values by voting against the Party with greater olig, although Voters might have been able to prevent the Party with greater olig from winning elections and setting the tax rate by its olig position. Similarly, Oligarchs were not able to alter the mean olig value by donating to the campaign of the Party with greater olig, although Oligarchs might have been able to increase their favored Party's chance of winning elections and setting the tax rate.

A one-tailed Welch's two-sample t-test confirmed our expectation that Oligarchs would donate less money on average when Party olig values were fixed compared to in the default model, $p < 2.2e-16$. The mean fraction of Oligarch gross-income donated in the base model was 0.112, and in the fixed Party olig model it was only 0.092, or less than one tenth as much (Figure 14). Due to a high variance in donation level, however, Cohen's d for the difference in donation levels between the two models is only a small 0.26. Note that Figure 14 and all other box plots in this work are Tukey box plots, where the middle line is the median, the lower and upper box ends are the lower and upper quartiles, the lower staple marks the lowest datum within 1.5 times the box height of the lower quartile, and the upper staple marks the highest datum within 1.5 times the box height of the upper quartile.

Another one-tailed Welch's two-sample t-test showed that Oligarchs earned higher profits in the fixed Party olig model than in the base model, $p < 2.2e-16$. Oligarchs in the fixed Party olig model earned approximately 2.38 times the profits of Oligarchs in the base model on average, for a very large Cohen's d of 7.04. Oligarchs' profits were higher because they were spending much less on campaign donations, and the expected Party olig value of zero at initialization was greater than the mean Party olig value of -45.3 from the default model, so taxes tended to be higher in the fixed Party olig model.

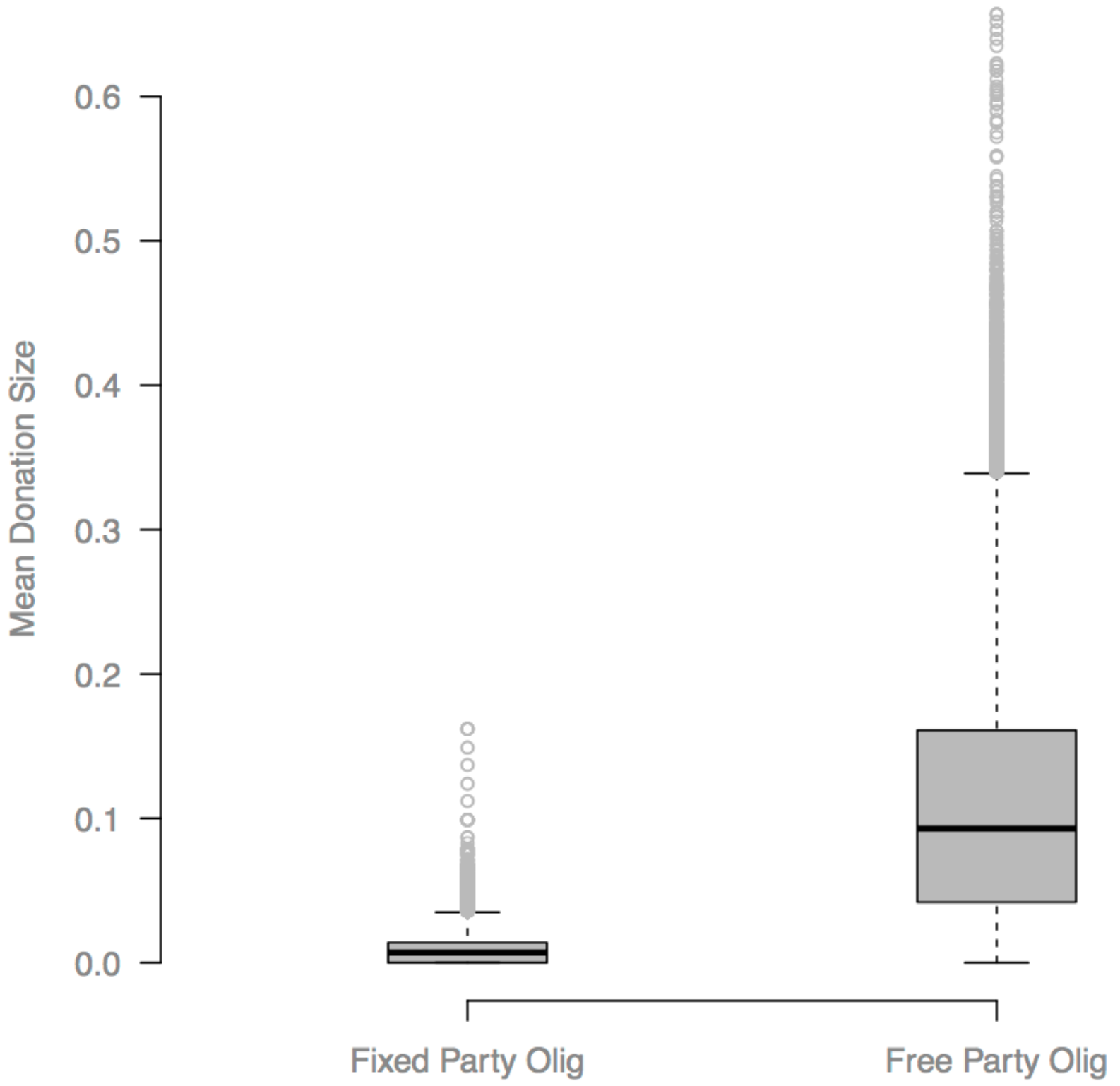

*Figure 14*. Mean fraction of gross-income Oligarchs donated. Each time step of each run produced one data point.

## Results: Voter olig-salience parameter sweep

In the Voter olig-salience parameter sweep, Voters were initialized to all have the same olig-salience value, ranging from 0 to 1.0 in intervals of 0.2 over the sweep. Voters were not allowed to change their olig-salience values during a model run. This meant that they did not adapt to different tax rates by changing their preferences over Party positions.

We hypothesized that when olig-salience is higher, Oligarchs would have less success in influencing policy. This would be evident in the form of the following outcomes: lower tax rates, lower mean Party olig value, lower Oligarch profits, and a lower frequency of the Party with greater olig value defeating an opponent with a more centrist ideo value. Because Oligarch campaign donations would be less likely to increase tax rates with a high Voter olig-salience, Oligarch donations would be lower when olig-salience is fixed at higher values. The downside for Voters of a high Voter olig-salience, we hypothesized, was that it would allow the mean Party ideo position to be farther from zero, the expected preference of the median Voter.

Our hypotheses were supported by the results (Table 3). There was a strong negative Spearman correlation (-0.76) between olig-salience and the frequency of the Party with greater olig defeating a more centrist Party, and there was a moderate positive correlation (0.32) of olig-salience with the absolute value of the mean Party's ideo value (Figure 15).

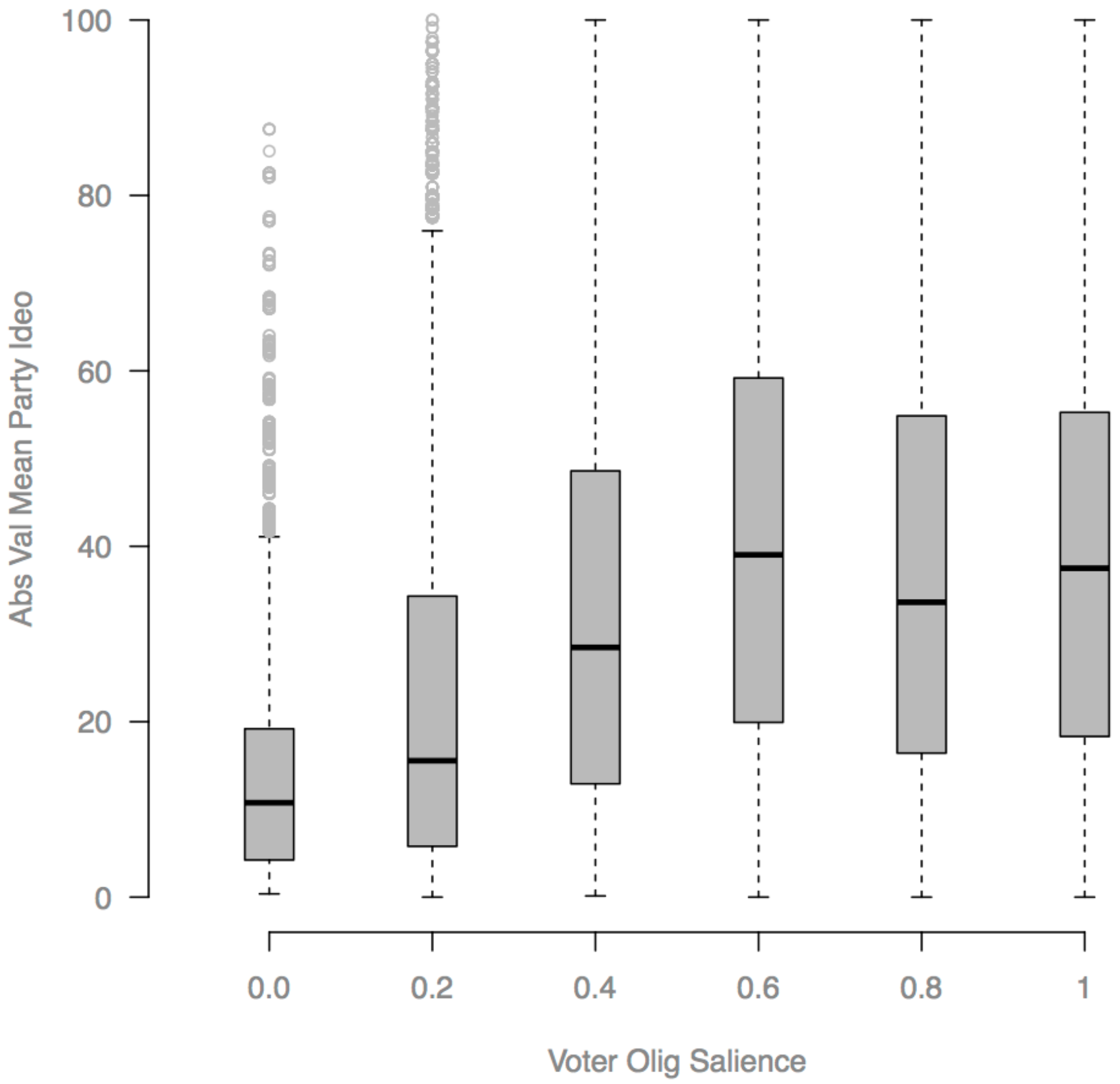

*Figure 15*. Absolute value of the mean Party ideo position, from the Voter olig-salience parameter sweep. Each time step of each run produced one data point. Spearman correlation = 0.32.

For the other outcome variables, however, correlations were weak until the two highest olig-salience parameter settings (0.8 and 1.0) were dropped. This is because when the olig-salience value is fixed at a very high value, the OLIGO model degenerates into a one-dimensional policy space model, where Voters all share a common position at one extreme of the sole policy axis. In this degenerate case, the Hunter strategy that the Parties follow does not work well, because if the Parties are not within a step length of each other along the one meaningful policy axis, the losing Party cannot improve its vote count no matter which direction it moves, because it will still be farther from all Voters. Therefore, both Par-ties will wander aimlessly, until the losing Party happens to cross paths with the winning Party. As a result, when olig-salience values are greater than about 0.6, Oligarch power is likely to increase. This degenerate phenomenon is not usually a problem for the OLIGO model, because such high mean Voter olig-salience values almost never occur, except in the parameter sweep.

| Parameter sweep | Profit corr. | Donation corr. | Tax corr. | Olig-defeats-center corr. |
|---|---|---|---|---|
| Ad-decay-sweep | -0.27 | -0.26 | -0.28 | -0.76 |
| Voter-memory-sweep | -0.37 | -0.14 | -0.37 | -0.37 |
| Olig-salience-sweep | -0.21 | -0.17 | -0.21 | -0.76 |
| Olig-salience-sweep low-sal | -0.57 | -0.35 | -0.59 | -0.81 |
| Donation-size-sweep | 0.16 | . | 0.26 | 0.86 |

*Table 3*. Correlations of sweep parameter with mean Oligarch profit, mean Oligarch donation parameter, mean tax rate, and fraction of runs where Party of higher olig position defeats Party of more centrist ideo position. Olig-salience-sweep low-sal gives results for the olig-salience sweep, omitting the conditions with olig-salience equal to 0.8 or 1.0. Donation level correlation is not reported for the donation-size-sweep because donation level is the sweeping parameter.

When the 0.8 and 1.0 olig-salience parameter conditions were dropped, leaving only the conditions (0, 0.2, 0.4, 0.6), meaningful trends emerged. There was a moderate to strong negative correlation between olig-salience and tax rate (-0.59) mean Oligarch profit (-0.57), mean Party olig (-0.54), and mean Oligarch donation fraction (-0.35). A box plot of the effect of olig-salience on tax rates showed a clear negative trend up to an olig-salience of 0.6, followed by an upward trend for very high olig-salience values (Figure 16).

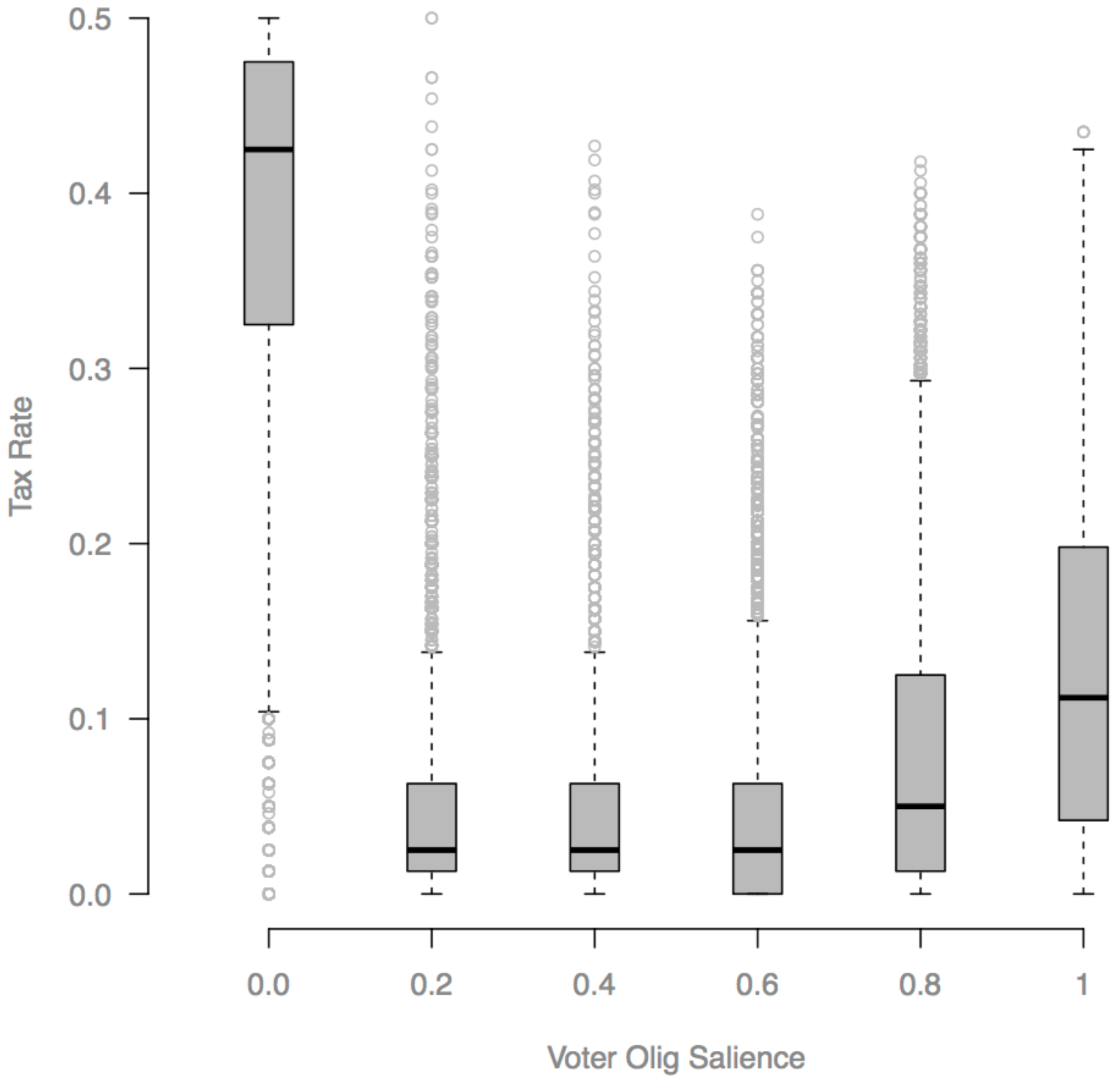

*Figure 16*. Tax rate from the Voter olig-salience parameter sweep. Each time step of each run produced one data point. Spearman correlation (up through olig-salience 0.6) = -0.59.

## Results: Oligarch donation-size parameter sweep

In the Oligarch donation-size parameter sweep, the donation-size of all Oligarchs was fixed to a certain value at the beginning of each run, and Oligarchs were not allowed to change this value. Recall that the donation-size parameter sets the maximum fraction of gross-income that an Oligarch can donate. The actual fraction of gross-income donated by an Oligarch is the product of the donation-size parameter and a scaling factor that increases with the difference between the Parties' olig positions.

We hypothesized that in the donation-size parameter sweep, greater donation-size values would produce better results for Oligarchs, up until the highest donation-size values, where diminishing returns would make profits decline.

Our hypotheses were mainly supported by the results. There were moderate to strong positive correlations between donation-size and the frequency of a higher olig Party defeating a more centrist ideo Party (0.86) and mean Party olig value (0.38), and a weak positive correlation of donation-size with tax rate (0.26). Oligarch profits had a positive correlation with donation-size, but at 0.16, it was too low to be meaningful. The box plot in Figure 17 shows that there was a clear positive trend in profits with increasing donation-size, although past 0.6, variance increased and losses became common.

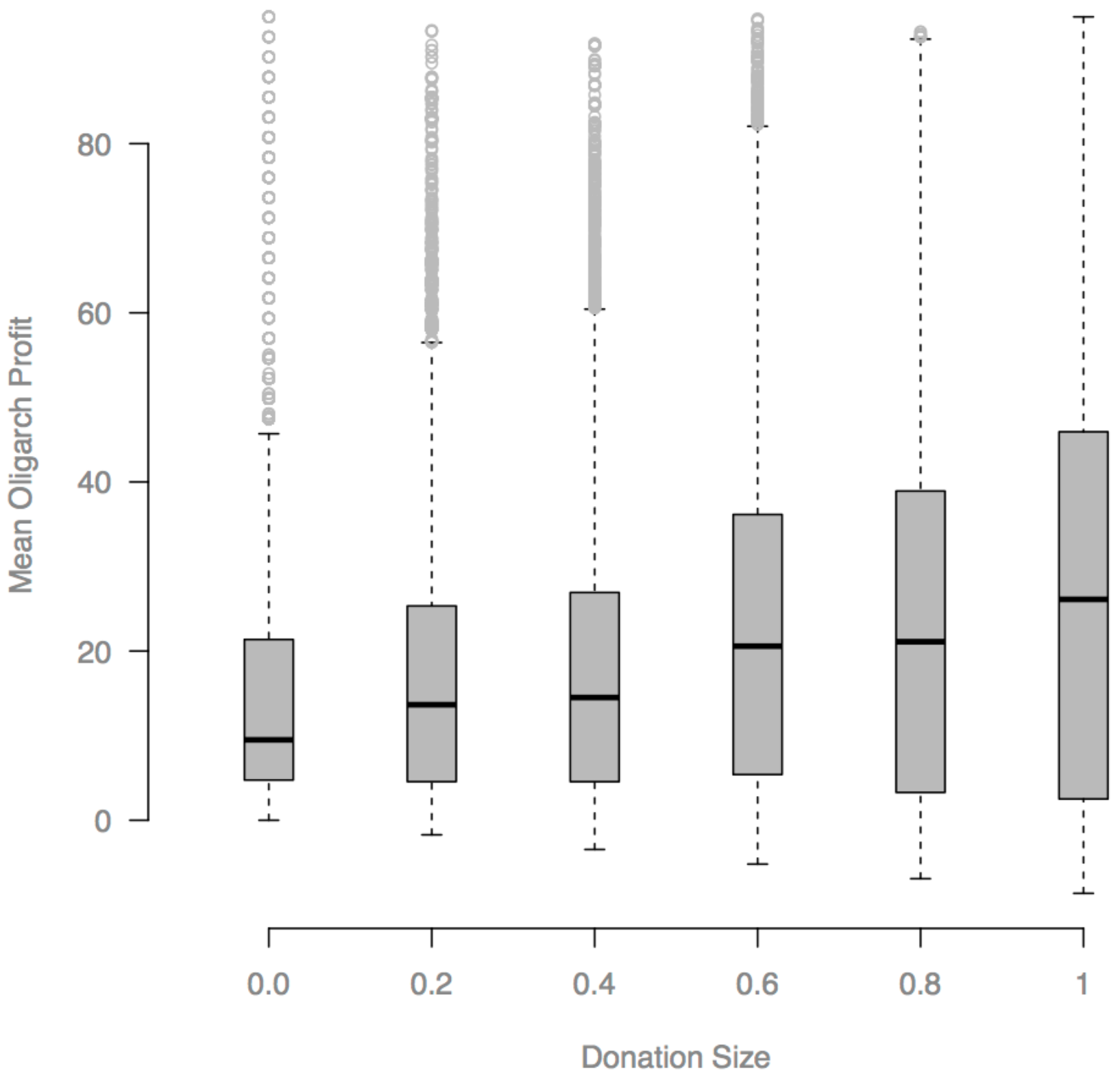

*Figure 17*. Mean Oligarch profit from the Oligarch donation-size sweep. Each time step of each run produced one data point. Spearman correlation = 0.16.

## Results: Voter awareness sweeps

Two parameter sweep experiments tested the effect on model outcomes of settings that adjust Voter awareness of tax rates. In the ad decay factor sweep, the VOTER-AD-DECAY-FACTOR parameter was set in the range from -0.12 to -0.02, in steps of size 0.02. Recall that the VOTER-AD-DECAY-FACTOR determines the effect of a campaign donation on the distance Voters perceive to a Party. A higher ad decay factor (less negative) causes a given campaign donation to have more influence on Voter perceptions.

In the memory strength sweep, the VOTER-MEMORY-STRENGTH parameter was adjusted from 0.1 to 0.9, in steps of size 0.2. The VOTER-MEMORY-STRENGTH parameter is the forgetting rate, that is, the factor by which a Voter's olig-salience is multiplied at each run step if the Voter does not "notice" a high tax rate. The higher the memory strength, the more slowly a Voter's olig-salience will decline between events that cause it to increase.

We predicted that increasing ad decay factor, and increasing voter memory strength, would each be correlated negatively with mean Oligarch donation-size, and with measures of Oligarch success, including the frequency of the Party with higher olig value defeating a Party of more centrist ideo value, tax rate, mean Oligarch profit, and mean Party olig position.

The results confirmed our expectations. In the ad decay sweep, we found moderate to strong negative correlations of VOTER-AD-DECAY-FACTOR with the frequency of a higher olig Party defeating a centrist Party (-0.76), tax rate (-0.28) (Figure 18), mean Oligarch profit (-0.27), mean Party olig value (-0.26), and mean Oligarch donation-size (-0.26) (Figure 19). Similarly, in the memory strength sweep, we found negative correlations of VOTER-MEMORY-STRENGTH with the frequency of a higher olig Party defeating a centrist Party (-0.37), tax rate (-0.37), mean Oligarch profit (-0.37), mean Party olig value (-0.42), and mean Oligarch donation-size (-0.14).

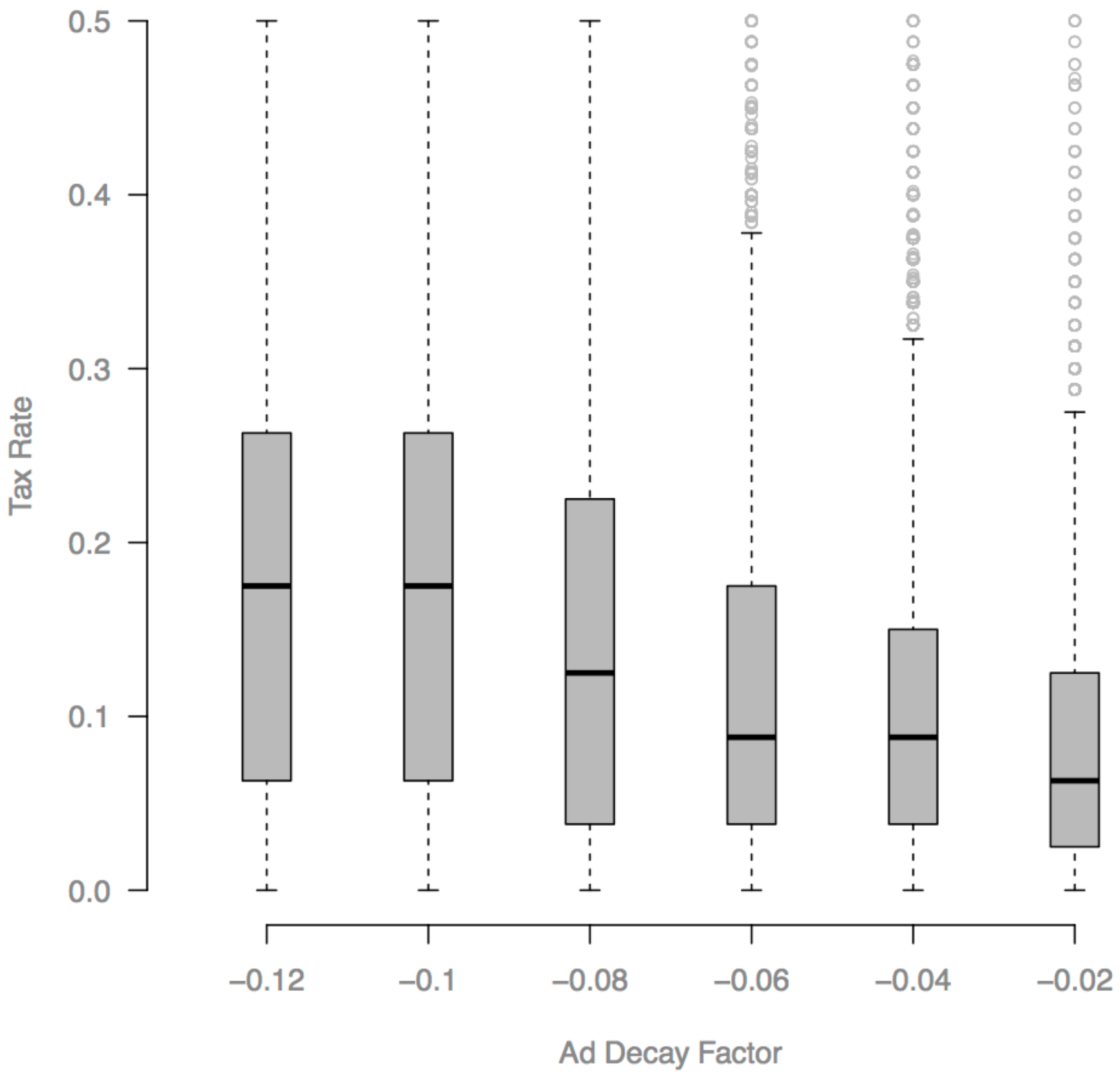

*Figure 18*. Tax rate from the ad decay sweep. Each time step of each run produced one data point. Spearman correlation = -0.28.

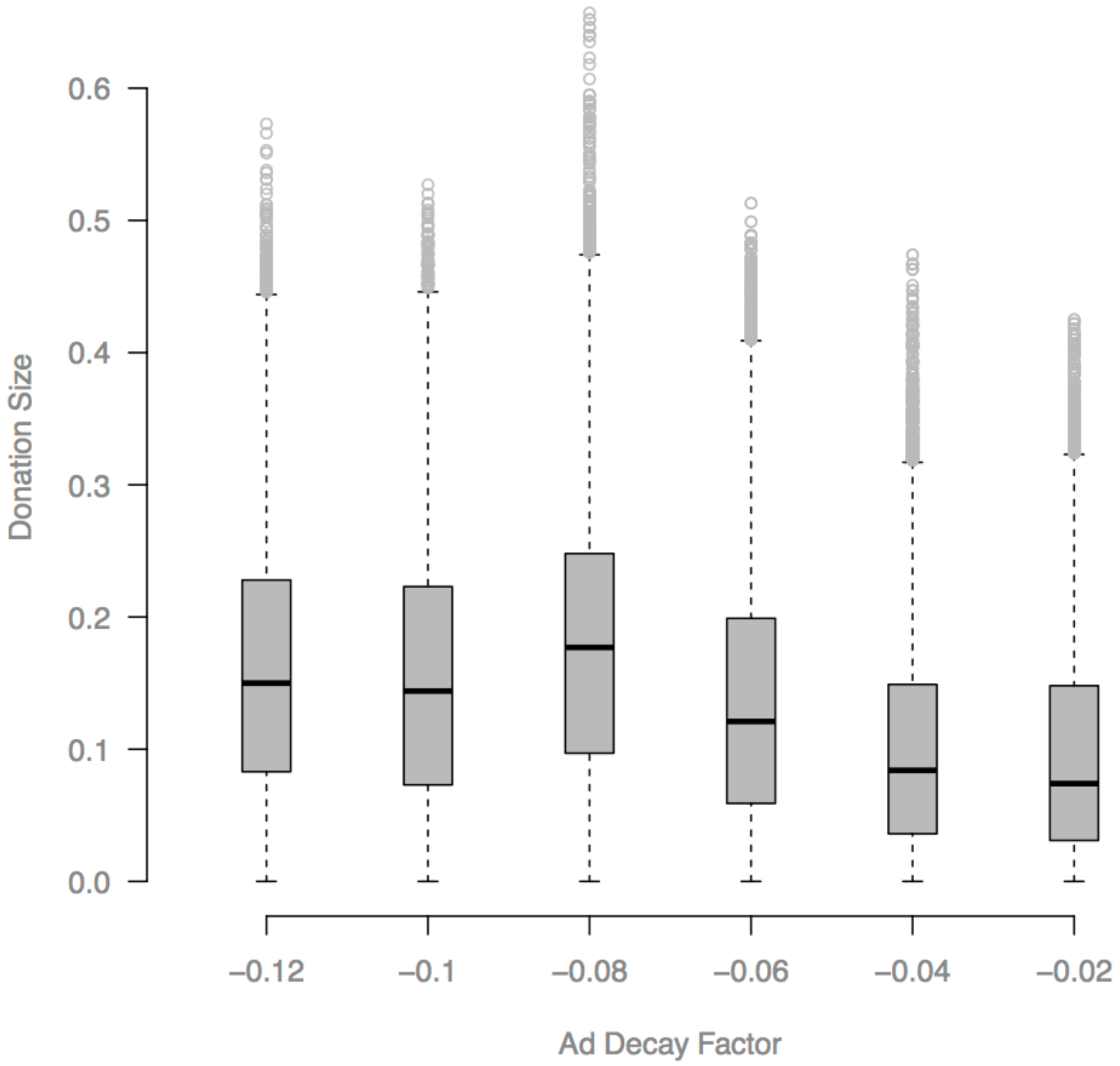

*Figure 19*. Mean Oligarch donation-size from the ad decay sweep. Each time step of each run produced one data point. Spearman correlation = -0.26.

## Results: Party-Ideo-Motivated Model

In the Party-Ideo-Motivated Model (PIMM), Parties move along the ideo axis by adding a bias term to the position they would ordinarily take as they adapt to the reward-and-punishment of polling results. This bias term tends to steer Parties toward their preferred ideo positions of -33 (blue Party) and +33 (red Party). Results indicate that the mean ideo position for the red Party in the PIMM is 5.2, versus -5.2 for the blue Party, compared with 0.0 for both Parties in the IIM (Cohen's d = 1.06). In the PIMM as in the IIM, Voters' ideo preferences have a mean of approximately 0, which pushes Parties toward centrist ideo policies, but the Parties' own biases in the PIMM push them back toward their respective bliss points.

The correlation between Oligarch donations and the frequency with which the Party with greater olig value defeats a more centrist Party is much weaker in PIMM compared to IIM: The correlation is 0.353 in PIMM versus 0.864 in IIM, and the difference is highly significant ($p < 0.001$) using Fisher's r-to-z and a t-test. This result makes sense, because the loyalist Voters in the PIMM dampen the effect of Oligarch donations to the Party with higher olig position. Loyalist Voters are not affected by campaign donations and vote based only on Party ideo position, so they tend to support centrist Parties rather than Parties that receive high donations.

The correlation between mean Oligarch profit and Voter olig-salience is greater in PIMM than in IIM: The correlation is 0.716 in PIMM versus 0.678 in IIM, and the comparison is statistically significant (p = 0.099) using Fisher's r-to-z and a t-test. The correlation between Oligarch profit and Voter olig-salience may be more pronounced in PIMM because loyalist Voters in PIMM do not change their votes as a result of high taxes, so salience must become greater among swing voters before they can vote out a Party that has a high olig position.

## Results: Validation of the PIMM with Polling Data

We performed experiments to validate the performance of the Party-Ideo-Motivated Model (PIMM) against real world data on changes over time in two-party opinion polls. As outcome measures for comparison with opinion polls, we analyzed the mean absolute change in percent support for the conservative Party, the minimum observed percent support for the conservative Party, the standard deviation of change in percent support for the conservative Party, and the maximum observed absolute change in percent support for the conservative Party. That is, we investigated the following questions: How well does our model reflect rates of change in party support over time, from real world polling data? For example, does our model reflect the typical monthly percent change in support for a party in the US two-party system? Does the range of observed support levels for a party reflect the observed range over a corresponding time period in US polling data? Is the range of percent changes in support for a party similar to that in the US system? All observations were over periods of 120 cycles, with 100 trials each for the IIM, PIMM, and AIMM. The size of differences from the observed polling results was evaluated via Cohen's d.

The PIMM was only trivially different from polling results in mean absolute change in percent support (Cohen's d = 0.001). See Table 4. PIMM had a slightly higher minimum observed conservative percent support than the polls (d = 0.266). PIMM did, however, have a moderate increase in standard deviation of change in percent support from 2.87 to 3.31 (d = 0.666), and a large increase in maximum observed absolute change in percent support from 8 to 11.57 (d = 1.266). Even so, PIMM appears to provide a reasonable approximation to the changes over time observed in percent support for two parties in the polling data.

| Model | Mean red vote | Max red vote | Mean absval red vote change | Max absval red vote change | Within-run $\sigma$ red vote change |
|---|---|---|---|---|---|
| Gallup | 42. | 50. | 2.28 | 8. | 2.87 |
| IIM | 48.82 | 81.21 | 8.09 | 51.74 | 13.01 |
| PIMM | 50.87 | 65.55 | 2.28 | 11.57 | 3.31 |
| AIMM | 50.3 | 74.4 | 2.25 | 17.2 | 3.38 |

*Table 4*. Comparison of Gallup poll, Ideo-Indifferent Model, Party-Ideo-Motivated Model, and All-Ideo-Motivated Model results.

The basic model version, IIM, had large differences from the polling data across all measures. IIM had much lower minimum observed conservative percent support, 17.6 compared to 34 in the polls (d = 1.448). IIM also had much higher mean absolute change in percent support, at 8.1 rather than 2.3 (d = 2.961). Maximum observed absolute change in percent support was much higher in IIM than in the polls, at 51.7 versus 8 (d = 2.514). The standard deviation of change in percent support is higher in IIM, at 13.00 versus 2.87 (d = 2.526). In general, IIM has much greater variability in party support from one step to the next than either PIMM or the poll results. IIM also produces a much greater range of party support percentages than are observed in the real world data. (The All-Ideo-Motivated Model, introduced below, produced results somewhere between the large jumps in votes of the IIM and the more realistic, smaller jumps of the PIMM.)

These results are reasonable, as they show that the introduction of loyalist Voters and Party ideological preferences in the Party-Ideo-Motivated Model leads to reduced variance and range in percent Party support, compared to the Ideo-Indifferent Model. Two-thirds of Voters in the PIMM are loyalists, i.e., they vote based on the ideo axis positions of the two Parties and are not swayed by changes in tax rate, Party olig position, or campaign donations. The presence of loyalist Voters dampens the effects of changes in tax rate, Party olig position, and Oligarch donation level. Changes in party support from one step to the next tend to be decreased. The result is changes in party support over time that more closely reflect real world polling data in the PIMM, compared to the IIM.

We focus in our reporting of results on the Ideo-Indifferent Model outcomes, because the IIM is simpler to describe than the PIMM, it still captures key features of our model, and its results are very similar to the results of the PIMM, except in terms of the rate of change in party support. We do, however, present results from the PIMM and AIMM for comparison in table form.

## Results: All-Ideo-Motivated Model

Recall that the All-Ideo-Motivated Model (AIMM) is similar to the Party-Ideo-Motivated Model, but in the AIMM, Oligarchs have a preferred Party ideo position of +33. The Oligarch's preferred ideo position is also the same as the conservative Party's bliss point on the ideo axis. Oligarchs in the AIMM choose which Party to donate to based on a combination of Party ideo position and olig position, unlike in the IMM and PIMM, in which Oligarchs are indifferent to Party ideo position.

We will discuss AIMM results by comparing them to PIMM results. In the AIMM, Oligarchs are less efficient at capturing high tax rates than in the PIMM, and also donate less in the AIMM (Table 2). In the AIMM, mean Oligarch donations are much lower than in PIMM, at 0.06 versus 0.10 (Cohen's d = 1.28). The mean olig-salience of Voters is much lower in AIMM, at 0.067 versus 0.083 (d = 0.89). The mean tax rate is also lower in AIMM, at 0.085 versus 0.092 (d = 0.34). Olig-salience, tax rate, and donation level are all measures of the size and effi-ciency of Oligarch influence, and all indicate lower influence in AIMM. Note that similar to IMM and PIMM, in AIMM, Oligarchs adjust their donation levels based on the same UPDATE-OLIGARCH-DONATIONS function. This means that Oligarchs will decrease their donation levels under two conditions: (1) if their profits decreased after an increase in donations, or (2) if profits increased after a decrease in donations. This also implies that when Oligarchs in AIMM donate to a Party with greater ideo position but lower olig position, they donate "with their hearts" – i.e., these donations are ideologically motivated, and are not based on their profit motive. If the Party with lower olig position wins, the Oligarchs' profits will be lower than if Oligarchs had not donated. Therefore, the likely reason why Oligarchs tend to donate less in AIMM is that their donations sometimes reduce profits.

In AIMM, the conservative Party is ideologically preferred by Oligarchs. Our experiments show that the conservative Party tends to adopt a slightly *lower* olig value (i.e., an olig value *less* preferred by Oligarchs) compared to its rival Party. In other words, the Party that Oligarchs naturally favor ideologically tends to adopt a tax policy on the olig axis that Voters prefer but Oligarchs do not. The mean olig position of the conservative Party in AIMM is lower than in PIMM, at -43.4 versus -39.7 (Cohen's d = 0.13). This leads to a slightly lower mean olig position in AIMM compared to PIMM, at -41.4 versus -39.7 (d = 0.12). The effect sizes of these trends are very small, but the trends hold consistently across trials. It appears that the Party naturally favored by Oligarchs takes advantage of its preferred position to court Voters, through a lower olig position. Oligarchs may donate to the conservative Party in spite of its lower proposed tax, due to their preference for conservative ideo positions.

# Summary and Discussion

To summarize, our simulation experiments produced the following findings:

- Result 1. Oligarchs tend to donate less to political campaigns when Parties' proposed policies are inflexible, when Voters are less susceptible to advertising, or when Voters are strongly opposed to Oligarch interests
- Result 2. When Voters' awareness of subsidies to Oligarchs is greater, taxes on Voters tend to be lower
- Result 3. Greater Oligarch campaign donations produce higher subsidies to Oligarchs, but may produce diminishing Oligarch profits if donations are too high
- Result 4. If oligarchs choose which party to support based on ideological goals instead of expected profits, and adjust their donation levels based strictly on subsequent profits, then equilibrium donation levels will be lower.

We discuss these results next.

Result 1 can be understood as follows: Under certain conditions (i.e., when Parties' proposed policies are inflexible; when Voters are less susceptible to advertising; or, when Voters are strongly opposed to Oligarch interests), Oligarchs' marginal benefit from increased donations is lower than usual, while their marginal cost is the same as always. Hence, Oligarchs' marginal cost will equal their expected marginal benefit from increased donations at a lower donation level than usual. Above that donation level, Oligarchs will tend to lose money if they increase their donations, so their Hunter strategy will typically cause them to decrease their donations in the next cycle after such an increase. This result supports Brock and Magee (1978) and Bardhan and Mukherjee (2000), who argued that Oligarchs can profitably donate money to the campaign of a Party that favors their economic interests until the marginal cost of donating more becomes greater than the marginal benefit of an increased donation.

Result 2 can be understood as follows: When Voters attach high salience to the olig policy axis, they are more likely than normal to vote for the Party with lower olig position. As a result, Parties can gain votes more reliably by adopting low olig positions, so their Hunter strategy will tend to direct Parties toward low olig values. The incumbent Party will typically have a low olig value, which produces lower tax rates. We also found that greater Oligarch campaign donations in our model tend to produce higher subsidies, but they can also produce diminishing Oligarch profits if donation levels become too high (Result 3). These results support findings by Grossman and Helpman (1996), who suggested that the more strongly Voters are opposed to a policy, the more expensive it is to campaign successfully to elect representatives who endorse the policy. According to Grossman and Helpman, if a politician has the option to adopt a policy favored by potential donors but opposed by voters, the politician will not adopt the policy unless the increased campaign donations are enough to bring at least as many votes as would be lost in moving to the less popular position.

We used the Party-Ideo-Motivated Model to test the effects of giving each Party a preferred ideological position, and of making some Voters "loyalists" who vote based on ideology and ignore the *olig* axis. This model produced a better fit to Gallup polling data on Democratic and Republican Party support over time than the base model (IIM). The PIMM results show reduced impact of donation levels on election outcomes, due to the presence of loyalist voters who are unaffected by campaign donations. Otherwise, the results support the conclusions drawn from the IIM, though with slightly blunted effect sizes.

With the All-Ideo-Motivated Model (AIMM), we analyzed the effect of assigning oligarchs an ideological preference, keeping all else the same as the PIMM. We discovered a novel result (Result 4), that when oligarchs donate to parties based on a combination of ideological and profit motivations, adjusting their donation levels based on the resulting profits, oligarchs tend to reach a lower equilibrium donation level, because the lost profits under this donation policy discourage oligarchs from continuing to make large donations. In future work, we envision modeling the likely effect of incumbent party ideology on oligarch profits.

It would be interesting to see whether oligarchs would still arrive at a lower equilibrium level of donations in AIMM, if they adjusted donation levels based partly on party ideology, or if incumbent party ideology influenced oligarch profits.

To conclude, we remind readers of the pragmatic motivation for our work. Special interest groups now exert considerable influence on US elections and lawmaking through campaign donations and lobbying, particularly since the rise of PACs. Pressure groups spend $2 billion annually on lobbying the US federal government and $500 million per election cycle in campaign spending (Baumgartner, Berry, Hojnacki, Leech, & Kimball 2009). Recently, interest groups have fought successfully against return-free tax filing, direct-to-consumer auto sales, and low limits on H-1B visas (Day 2013; Salustri 2014; LaPlante 2007). The OLIGO model is a novel agent-based model that captures key features of prior theory on special interest campaign donations. Our results suggest that well-informed voters, uncompromising party leaders, and limited campaign donations can reduce the influence of special interest groups in a democracy, and furthermore, under such conditions, the wishes of the median voter may have a greater impact on legislative action.

# Acknowledgements

The research was supported in part by an NSF Early CAREER Award (OCI 1150230; PI: Pratim Sengupta). All opinions presented herein are the authors', and are not endorsed by any supporting institution. The authors gratefully acknowledge Andrew Hostetler and Rich Lehrer for their support during this work.

# Appendix: OLIGO model details

## Entities, State Variables, and Scales

Global parameters of the OLIGO model are included here for completeness, but unless otherwise specified, these settings can be changed without substantially altering the results of a model run. See Figure A1. The GDP, or total income of all agents in the model, is fixed at 1000.

Oligarchs adjust the fraction of their gross-income that they donate to a Party by step size (OLIGARCH-EPSILON) of 0.1, or 10%. Oligarchs collectively earn 0.05 of model GDP (OLIGARCH-GDP-FRACTION), as their total gross-income. Oligarchs are initialized to donate a random amount, up to 0.3 (OLIGARCH-INITIAL-DONATION), of their gross-income. Oligarchs will choose to increase or decrease their donation-size when they would otherwise do the opposite, with probability 0.1 (OLIGARCH-NOISE), to allow them to avoid becoming trapped in suboptimal behaviors.

Parties adjust their positions along the ideo and olig axes with a step size of 5 (PARTY-EPSILON), in the range of -100 (MIN-POS) to 100 (MAX-POS). The PARTY-DIFFERENCE-FACTOR affects Oligarchs' perception of how different the Parties are along the olig policy axis. If the Parties are close together along the olig axis, Oligarchs will donate less to their favored Party.

| Parameter | Value |
|---|---|
| GDP | 1000. |
| MAX-POS | 100. |
| MIN-POS | -100. |
| OLIGARCH-COUNT | 5. |
| VOTER-COUNT | 100. |
| OLIGARCH-EPSILON | 0.1 |
| OLIGARCH-GDP-FRACTION | 0.05 |
| OLIGARCH-INITIAL-DONATION | 0.3 |
| OLIGARCH-LATENCY | 10. |
| OLIGARCH-NOISE | 0.1 |
| PARTY-DIFFERENCE-FACTOR | -2. |
| PARTY-EPSILON | 5. |
| VOTER-AD-DECAY-FACTOR | -0.03 |
| VOTER-AWARENESS | 0.5 |
| VOTER-GROSS-INCOME | 9.5 |
| VOTER-INITIAL-SALIENCE | 0. |
| VOTER-MAX-TAX | 0.5 |
| VOTER-MEMORY-STRENGTH | 0.9 |
| VOTER-MIN-DISTANCE-SCALE | 0.3 |
| VOTER-OLIG | -100. |

*Figure A1.* Global parameters.

## Initialization of Oligarchs

Because Oligarchs' total gross-income is intended to be a fixed fraction of the total gross-income of the population, Oligarchs' incomes are adjusted by the difference between their sample mean and the desired mean, in a first moment matching procedure. If any Oligarch's income becomes negative during moment matching, all Oligarchs' gross-income values are drawn again until moment matching can be completed (Figure A2). Each Oligarch's donation-size value is set to a random value up to the parameter OLIGARCH-INITIAL-DONATION.

```
INITIALIZE-OLIGARCHS:
    target-mean ← GDP * OLIGARCH-GDP-FRACTION /
        OLIGARCH-COUNT
    target-standard-dev ← target-mean / 2
    target-var ← target-standard-dev^2
    μ ← ln (target-mean^2 / sqrt (target-var + target-mean^2))
    σ ← sqrt ln(1 + target-var / (target-mean^2))
    for i ← 1 to OLIGARCH-COUNT
        create Oligarch
            gross-income ← RANDOM-LOGNORMAL(μ, σ)
                donation-size    ←    RANDOM-FLOAT(OLIGARCH-INITIAL-
DONATION)
    target-mean ← GDP*OLIGARCH-GDP-FRACTION/OLIGARCH-COUNT
    mean-oligarch-income ← mean gross-income of Oligarchs
    for Oligarchs
        gross-income ← gross-income − (mean-oligarch-income − target-mean)
    if any Oligarch has gross-income < 0
        delete Oligarchs
        INITIALIZE-OLIGARCHS
```

*Figure A2.* Oligarch initialization submodel.

# Submodels

The first submodel enacted in each run step is UPDATE-OLIGARCH-DONATIONS (Figure A3), in which Oligarchs adjust their donation-size values. To prevent Oligarchs from changing their donations in lockstep, an OLIGARCH-LATENCY parameter of 10 cycles is used, which means that each Oligarch updates its donation-size only every 10 run cycles, and no two Oligarchs will update their parameter in the same cycle in a population with fewer than 11 Oligarchs. If an Oligarch's expected profit has increased since 10 cycles earlier, when the Oligarch most recently changed its donation-size, the Oligarch will change its donation-size again in the same direction; otherwise, the Oligarch will change its donation-size in the opposite direction. The Oligarchs' strategy is based on the Hunter behavior proposed by Laver (2005). To allow Oligarchs to explore alternative strategies, Oligarchs are induced to make a change in the opposite direction from what they would do otherwise, with probability OLIGARCH-NOISE. The actual amount an Oligarch donates to the Party with greater olig value is the Oligarch's donation-size parameter, multiplied by the Oligarch's gross-income, multiplied by a scaling factor that increases with the distance between the two Parties' olig values. See Figure A4 for the calculation of the scaling factor. If the two Parties' olig values are similar, an Oligarch will donate less money to its preferred Party, because it can expect less benefit from donating, as suggested by Brock and Magee (1978).

We tested a variation on this model where Oligarchs had a third option, the "null action," in which they did not change their donation level from the previous cycle. Oligarchs would change their donation level only if the reward or punishment in changed profit of the previous action exceeded a threshold value, otherwise they would keep their donation-size parameter the same. This "null action" model variation led to only a trivial difference in outcomes, as will be shown in the results section.

```
UPDATE-OLIGARCH-DONATIONS:
    total-donations ← 0
    for Oligarchs
        if (cycle + agent-index) mod OLIGARCH-LATENCY == 0
            will-increase ← EXPECTED-DONATION-INCREASED ==
                EXPECTED-PROFIT-INCREASED
            if RANDOM-FLOAT(1) < OLIGARCH-NOISE
                will-increase ← ¬ will-increase
            if will-increase
                donation-size ← min(1, donation-size + OLIGARCH-EPSILON)
            else
                donation-size ← max(0, donation-size − OLIGARCH-EPSILON)
        total-donations ← total-donations + gross-income ∗ donation-size ∗
            PARTY-OLIG-DIFFERENCE-SCALED
    oligarch-total-donations ← total-donations
```

*Figure A3.* UPDATE-OLIGARCH-DONATIONS submodel.

```
PARTY-OLIG-DIFFERENCE-SCALED:
    party-olig-difference ← (max_Parties olig − min_Parties olig)/(MAX-POS−
        MIN-POS)
    return 1 − e^(PARTY-DIFFERENCE-FACTOR ∗ party-olig-difference)
```

*Figure A4.* PARTY-OLIG-DIFFERENCE-SCALED function.

The second submodel to execute in each run cycle is the VOTE submodel (Figure A5), in which Voters cast votes for the Party that seems closer to them in policy space. The key to this submodel is the VOTER-DISTANCE function (Figure A6), which determines how far each Party appears to be in policy space from a Voter. The VOTER-DISTANCE function measures distance in multiple dimensions by summing the difference between vectors in each dimension, instead of by taking the Euclidean distance, because policy positions along each axis are not meaningfully related to each other geometrically (Blais, Nadeau, Gidengil, & Nevitte 2001). The policy difference along each axis is multiplied by the Voter's salience value for that axis, and the sum of these terms is taken. Finally, the result is adjusted according to the total donations to the Party's campaign, such that the Party will appear closer if its has received a large amount of campaign donations, down to a limit of VOTER-MIN-DISTANCE-SCALE times its "true distance." We take VOTER-MIN-DISTANCE-SCALE to be 0.3, following the example of Muis (2010), who used a similar procedure to model the effect of advertising on voter behavior. The marginal effect of campaign donations is modeled as decaying exponentially as the level of donations increases (Joseph 2006), and we set the rate of decay through the parameter VOTER-AD-DECAY-FACTOR, which determines how quickly the effect of greater donations plateaus to the limiting factor, VOTER-MIN-DISTANCE-SCALE.

```
VOTE:
    for Voters
        add vote to Party with minimum VOTER-DISTANCE(Party)
    make Party with more votes incumbent
```

*Figure A5*. VOTE submodel.

```
VOTER-DISTANCE(Party):
    ideo-distance ← |ideo − ideo of Party|
    olig-distance ← |olig − olig of Party|
    absolute-distance-with-salience ← ideo-distance * (1 - olig-salience) +
        olig-distance * olig-salience
    if Party has the higher olig
        return (VOTER-MIN-DISTANCE-SCALE +
            (1 - VOTER-MIN-DISTANCE-SCALE) *
            e ^(VOTER-AD-DECAY-FACTOR * oligarch-total-donations)) *
            absolute-distance-with-salience
    else
        return absolute-distance-with-salience
```

*Figure A6*. VOTER-DISTANCE function.

The third submodel to execute in each run cycle is REDISTRIBUTE (Figure A7), in which the new incumbent Party sets the tax rate, and Oligarchs receive a subsidy based on the amount of tax collected. The tax rate is derived from the incumbent Party olig value: It is set such that the fraction of the maximum allowable tax (0.5) equals the incumbent Party's fraction of the maximum allowable distance from the MIN-POS olig position (-100). All Voters have the same gross-income, so they all pay the same tax amount. Each Oligarch has a different gross-income, and each receives a share of the total subsidy in direct proportion to its fraction of the total gross-income of all the Oligarchs. Oligarchs receive subsidies in proportion to their incomes, because subsidies in the OLIGO model are meant to simulate the effects of legislation such as tariffs or farm subsidies, which yield greater payoffs to larger businesses.

```
REDISTRIBUTE:
    voter-tax-rate ← (VOTER-MAX-TAX / 2) *
        (1 + incumbent-olig / MAX-POS)
    tax-collected ← voter-tax-rate * VOTER-COUNT *
        VOTER-GROSS-INCOME
    for Oligarchs
        if (cycle + agent-index) mod OLIGARCH-LATENCY == 0
            last-subsidy ← tax-collected * gross-income /
                (GDP * OLIGARCH-GDP-FRACTION)
```

*Figure A7*. REDISTRIBUTE submodel.

The fourth submodel of each run cycle is UPDATE-VOTER-OLIG-SALIENCE (Figure A8). Voters adjust their olig-salience values in this submodel, in response to the current tax rate. Voters operate as reflex agents that may or may not recognize that they are paying a redistributive tax. The greater the tax rate, the more likely Voters are to receive the message and increase their olig-salience; this dynamic was suggested by Aidt (2000). We impose a limit on olig-salience, that it can be no greater than the tax rate. For example, if the current tax rate is $p$ in (0, 1), it is arguably not reasonable for a Voter to assign more than $p$ salience to the policy dimension that sets the tax rate, because tax rates are just one component of Voter utility, so all other components (modeled as the ideo dimension) should have a salience of at least $1 - p$. Two parameters help to determine a Voter's new olig-salience value. VOTER-AWARENESS determines how likely a Voter will be to increase its olig-salience at a given tax rate. A Voter becomes exponentially more likely to increase its olig-salience with a small increase in tax rates, if taxes are low. But if taxes are high, a small increase in tax rates will have less marginal effect on a Voter's likelihood of increasing its olig-salience. The second parameter is VOTER-MEMORY-STRENGTH, which determines how quickly a Voter "forgets" about past taxes over the course of successive run cycles. That is, this parameter determines how quickly the Voter's olig-salience decreases if the Voter does not notice new taxes. Our forgetting curve model of Voter behavior is modeled on the work of Aidt (2000).

```
UPDATE-VOTER-OLIG-SALIENCE:
  for Voters
    if RANDOM-FLOAT(1) < voter-tax-rate^VOTER-AWARENESS
      headroom ← min(0,voter-tax-rate - olig-salience *
        VOTER-MEMORY-STRENGTH)
      olig-salience ← olig-salience * VOTER-MEMORY-STRENGTH
        + RANDOM-FLOAT(headroom)
    else
      olig-salience ← olig-salience * VOTER-MEMORY-STRENGTH
```

*Figure A8*. UPDATE-VOTER-OLIG-SALIENCE submodel.

The last submodel to execute in each run cycle is UPDATE-PARTY-POLICIES (Figure A9), in which each Party moves to a new position in policy space. Each Party follows a Hunter strategy (Laver 2005), which means that it will move in the same direction as in the previous cycle if its votes increased, and otherwise it will move in a randomly chosen, different direction. Parties change their positions along each policy axis by the distance PARTY-EPSILON.

```
UPDATE-PARTY-POLICIES:
  for Parties
    if votes-increased
      increase-ideo ← last-increase-ideo
      increase-olig ← last-increase-olig
    else
      rand ← RANDOM-FLOAT(1)
        if rand < 0.5
          increase-ideo ← ¬ last-increase-ideo
          increase-olig ← ¬ last-increase-olig
        else if 0.5 < rand < 0.75
          increase-ideo ← ¬ last-increase-ideo
          increase-olig ← last-increase-olig
        else
          increase-ideo ← last-increase-ideo
          increase-olig ← ¬ last-increase-olig
    last-increase-ideo ← increase-ideo
    last-increase-olig ← increase-olig
    if increase-ideo
      ideo ← min(MAX-POS, ideo + PARTY-EPSILON)
    else
      ideo ← max(MIN-POS, ideo - PARTY-EPSILON)
    if increase-olig
      olig ← min(MAX-POS, olig + PARTY-EPSILON)
    else
      olig ← max(MIN-POS, olig - PARTY-EPSILON)
```

*Figure A9*. UPDATE-PARTY-POLICIES submodel.